\documentclass[11pt]{article}
\begin{document}
\title{Stiffness of polymer chains}

\author{A.D. Drozdov\footnote{
Phone: 972-86472146,
Fax: 972-86472916,
E-mail: aleksey@bgumail.bgu.ac.il}\\
Department of Chemical Engineering\\
Ben-Gurion University of the Negev\\
P.O. Box 653\\
Beer-Sheva 84105, Israel}
\date{}
\maketitle

\begin{abstract}
A formula is derived for stiffness of a polymer
chain in terms of the distribution function
of end-to-end vectors.
This relationship is applied to calculate the
stiffness of Gaussian chains (neutral and
carrying electric charges at the ends), chains modeled
as self-avoiding random walks, as well as
semi-flexible (worm-like and Dirac) chains.
The effects of persistence length and
Bjerrum's length on the chain stiffness
are analyzed numerically.
An explicit expression is developed for the radial
distribution function of a chain with the maximum
stiffness.
\end{abstract}
\vspace*{10 mm}


\section{Introduction}

This paper is concerned with stiffness of polymer
chains and explicit formulas for its determination.
As this issue lies on the border between statistical
physics of macromolecules and mechanical engineering,
it has not attracted substantial attention in the past.
The situation has changed dramatically in the past
decade due to the development of novel experimental
techniques for the direct measurements of force-extension
relations on individual chains in bio-polymers.
A correct measure of stiffness is important for our
understanding of the differences between the responses
of flexible and semi-flexible chains,
as well as for the assessment of the effects of segment
interactions (for example, excluded-volume interactions
for neutral macromolecules and electrostatic interactions
for polyelectrolyte chains) on their mechanical properties
(see \cite{CMM03} and the references therein).

In the statistical physics of polymers, the stiffness of a
macromolecule is conventionally associated with
the so-called ``effective spring constant:"
the ratio of a force applied to a free end of a chain
(its other end is assumed to be fixed)
to the displacement of the free end along the force
direction \cite{dGen79}.
An advantage of this method is that the stiffness
is determined in terms of the response of an individual
chain, which is convenient from the experimental standpoint.
A shortcoming of this approach is that it may lead to
conclusions that contradict the physical intuition.
As an example, we refer to the fact that this
``macro"-stiffness decreases with an increase in the
``micro"-stiffness characterized by the bending rigidity of
a semi-flexible chain.
A detailed discussion of this issue is provided in Section 2.

Mechanics of polymers focuses on the response of
ensembles of macromolecules and treats an elastic
modulus of an ensemble as a natural measure of
its stiffness.
A (shear or Young's) modulus is expressed in terms
of the strain energy density of a polymer,
and it reflects the cumulative contribution of
the strain energies of chains and the energies of
their interaction.
When the inter-chain interactions are negligible
(dilute polymer solutions) or their energy may be
adequately described in a simple way (the influence
of mutual interactions between chains in rubbery
polymers and polymer melts is traditionally accounted
for in terms of the incompressibility condition \cite{DE86}),
an elastic modulus of an ensemble divided by the
number of chains per unit volume provides a
natural measure of stiffness for an individual
macromolecule.

We apply the latter approach and derive an explicit
expression for the stiffness of a chain ``embedded"
into some ensemble subjected to affine deformations.
The objective of this study is to demonstrate that
the ``mechanical" definition of stiffness of a macromolecule
is free from the above shortcoming (an increase
in the bending rigidity of a semi-flexible chain
induces the growth of its stiffness, as it may be expected),
on the one hand,
and that it results in the same formula for the stiffness
of a Gaussian chain that the conventional definition
of stiffness does, on the other.
To develop an analytical formula for the stiffness,
we calculate the strain energy of an ensemble of chains
under an arbitrary affine deformation,
find the Young's modulus at uniaxial tension,
and associate the stiffness of an individual chain with
its contribution into the modulus.
This procedure allows an explicit expression to be
derived for the stiffness of a polymer chain with an
arbitrary end-to-end distribution function.
The formula is illustrated by several examples,
where the stiffness is expressed in terms of parameters
of the distribution function, and the dependence
of stiffness on these parameters is analyzed numerically.

The exposition is organized as follows.
In Section 2, the conventional stiffness of a chain
$\mu$ is calculated,
and it is found that $\mu$ is inversely proportional
to the mean square end-to-end distance of a chain.
The strain energy of an ensemble of polymer chains
with an arbitrary end-to-end distribution function
is determined in Section 3.
The formula for the strain energy is simplified
for uniaxial tension with small strains in Section 4.
An explicit expression for the stiffness of a chain
is derived in Section 5.
Section 6 focuses on several examples, where the
stiffness is determined analytically, and the effect
of parameters of the distribution function on
this quantity is studies numerically.
Some concluding remarks are formulated in Section 7.

\section{Conventional stiffness of a chain}

We begin with the conventional in statistical physics
definition of stiffness for a polymer chain.
Our aim is to develop an explicit expression for
this parameter and to demonstrate that it is inversely
proportional to $b^{-2}$, where $b$ is the mean square
end-to-end distance of the chain.

A chain is treated as a curve with some length $L$
in a three-dimensional space.
An arbitrary configuration of the chain
is described by the function ${\bf r}(s)$,
where ${\bf r}$ stands for the radius vector
and $s\in [0,L]$.
For definiteness, we assume the end $s=0$
to be fixed at the origin,
${\bf r}(0)={\bf 0}$, and the end $s=L$ to be
free.
An ``internal structure" of the chain is characterized
by a segment length $b_{0}$ and a number of
segments $N\gg 1$, which are connected with
$L$ by the formula $L=b_{0} N$.

A chain is entirely determined by its Hamiltonian $H({\bf r}(s))$.
Given $H$, the distribution of end-to-end vectors ${\bf Q}$
is described by the propagator (Green's function)
\begin{equation}
G({\bf Q})=\int_{{\bf r}(0)={\bf 0}}^{{\bf r}(L)={\bf Q}}
\exp \Bigl [-\frac{H({\bf r}(s))}{k_{\rm B}T}\Bigr ]
{\cal D}[{\bf r}(s)],
\end{equation}
where $k_{\rm B}$ is Boltmann's constant,
$T$ is the absolute temperature,
and the path integral is calculated over all curves
${\bf r}(s)$ that start at the origin and finish
at the point ${\bf Q}$,
\begin{equation}
{\bf r}(0)={\bf 0},
\qquad
{\bf r}(L)={\bf Q}.
\end{equation}
The path integral in Eq. (1) is determined unambiguously
by the normalization condition
\begin{equation}
\int G({\bf Q}) d{\bf Q}=1.
\end{equation}
For a chain loaded by a force ${\bf F}$ at the free end,
the Green function reads
\begin{equation}
G_{F}({\bf Q})=\int_{{\bf r}(0)={\bf 0}}^{{\bf r}(L)={\bf Q}}
\exp \Bigl [-\frac{1}{k_{\rm B}T}
\Bigl ( H({\bf r})-{\bf F}\cdot {\bf Q} \Bigr )
\Bigr ] {\cal D}[{\bf r}(s)],
\end{equation}
where the last term in the exponent stands for
the work of external force.
As this term is independent of the curve ${\bf r}(s)$,
Eqs. (1) and (4) imply that
\begin{equation}
G_{F}({\bf Q})=G({\bf Q})\exp \Bigl (
\frac{{\bf F}\cdot {\bf Q}}{k_{\rm B}T}\Bigr ) .
\end{equation}
Substitution of expression (5) into the partition function
$Z_{F}=\int G_{F}({\bf Q}) d{\bf Q}$
results in
\begin{equation}
Z_{F}=\int G({\bf Q})\exp \Bigl (
\frac{{\bf F}\cdot {\bf Q}}{k_{\rm B}T}\Bigr )d{\bf Q}.
\end{equation}
Differentiation of Eq. (6) with respect to ${\bf F}$
implies that
\begin{equation}
\langle {\bf Q}\rangle =\frac{k_{\rm B}T}{Z_{F}}
\frac{\partial Z_{F}}{\partial {\bf F}},
\end{equation}
where
$\langle {\bf Q}\rangle =\int {\bf Q} G_{F}({\bf Q}) d{\bf Q}
\Bigl [ \int G_{F}({\bf Q}) d{\bf Q} \Bigr ]^{-1}$
stands for the average end-to-end vector.
Formula (7) means that the vectors
$\langle {\bf Q}\rangle $ and ${\bf F}$ are
collinear, while their moduli are connected by
\begin{equation}
\langle Q\rangle =\frac{k_{\rm B}T}{Z_{F}}
\frac{\partial Z_{F}}{\partial F}.
\end{equation}
At small forces $F$, Eq. (8) can be linearized,
\begin{equation}
F=\mu \langle Q\rangle ,
\end{equation}
and the chain stiffness is described by the effective spring
constant $\mu$ in Eq. (9).

To reveal a disadvantage of this definition,
we consider the Marco--Siggia interpolation formula
\cite{MS95} for the force-stretch relation of
a semi-flexible (worm-like) chain,
\[
F=\frac{k_{\rm B}T}{l_{\rm p}}
\Bigl [ \frac{1}{4}\Bigl (1-\frac{u}{L}\Bigr )^{-2}
-\frac{1}{4}+\frac{u}{L}\Bigr ],
\]
where $l_{\rm p}$ stands for the persistence length.
It follows from this equality and Eq. (9) that
\begin{equation}
\mu=\frac{3k_{\rm B}T}{2l_{\rm p}L}.
\end{equation}
Bearing in mind that the persistence length $l_{\rm p}$
of a semi-flexible chain is connected with its bending
rigidity $\kappa$ by the formula
$l_{\rm p}=\kappa L/(k_{\rm B}T)$,
we find from Eq. (10) that the stiffness $\mu$ is
inversely proportional to the bending rigidity $\kappa$.
This conclusion appears to be counter-intuitive,
because one expects that the larger the rigidity
of a chain is at the micro-level (described by the
parameter $\kappa$), the higher its stiffness $\mu$
is at the macro-scale \cite{MKJ95}.

To show that this shortcoming is independent of
the concrete form of the force-stretch relation,
we derive an explicit expression for the pre-factor $\mu$
in Eq. (9) confining ourselves to chains with
isotropic Green functions $G=G_{\ast}(Q)$,
where $Q=|{\bf Q}|$.
Introducing a spherical coordinate frame
$\{ Q,\phi,\theta \}$ whose $z$ vector is directed
along ${\bf F}$, we find from Eq. (6) that
\[
Z_{F}=\int_{0}^{\infty} G_{\ast}(Q)Q^{2} dQ\int_{0}^{2\pi}d\phi
\int_{0}^{\pi} \exp \Bigl (\frac{FQ\cos\theta}{k_{\rm B}T}
\Bigr ) \sin \theta d\theta.
\]
Calculation of the integrals over $\phi$ and $\theta$
results in
\[
Z_{F}=4\pi \frac{k_{\rm B}T}{F} \int_{0}^{\infty}
G_{\ast}(Q)\sinh \Bigl (\frac{FQ}{k_{\rm B}T}\Bigr )Q dQ.
\]
Substitution of this expression into Eq. (8)
implies that
\[
\langle Q\rangle =-\frac{k_{\rm B}T}{F}
+\frac{\int_{0}^{\infty} G_{\ast}(Q)\cosh (\frac{FQ}{k_{\rm B}T})
Q^{2}dQ}{\int_{0}^{\infty} G_{\ast}(Q)\sinh (\frac{FQ}{k_{\rm B}T})
QdQ}.
\]
Setting $Q=b x$, where the mean square end-to-end distance
$b$ reads
\begin{equation}
b^{2}=\frac{\int_{0}^{\infty} G_{\ast}(Q)Q^{4}
d Q}{\int_{0}^{\infty} G_{\ast}(Q) Q^{2} dQ},
\end{equation}
and introducing the notation
\begin{equation}
f=\frac{F b}{k_{\rm B}T},
\qquad
u=\frac{\langle Q\rangle}{b},
\end{equation}
we arrive at the force-stretch relation
\begin{equation}
u=\frac{W_{2}(f)}{W_{1}(f)}-\frac{1}{f}
\end{equation}
with
\[
W_{1}(f)=\int_{0}^{\infty} G_{\ast}(b x)\sinh (fx)x dx,
\qquad
W_{2}(f)=\int_{0}^{\infty} G_{\ast}(b x)\cosh (fx)x^{2} dx .
\]
At small dimensionless forces $f$, the hyperbolic functions
are expanded into the Taylor series in $f$,
which implies that
\begin{eqnarray*}
W_{1}(f) &=& f\int_{0}^{\infty} G_{\ast}(b x)x^{2}dx
+\frac{f^{3}}{6} \int_{0}^{\infty} G_{\ast}(b x)x^{4}dx+\ldots,
\nonumber\\
W_{2}(f) &=& \int_{0}^{\infty} G_{\ast}(b x)x^{2}dx
+\frac{f^{2}}{2} \int_{0}^{\infty} G_{\ast}(b x)x^{4}dx+\ldots,
\end{eqnarray*}
where the dots stand for terms of higher order
of smallness.
Substituting these expressions into Eq. (13),
neglecting terms beyond the first order of smallness,
returning to the initial notation, and using Eq. (11),
we find that
\[
f=3u
\qquad
(f\ll 1).
\]
This equality together with Eq. (12) yields
\begin{equation}
F=\frac{3k_{\rm B}T}{b^{2}}\langle Q\rangle
\qquad
(\langle Q\rangle \ll b).
\end{equation}
Equations (9) and (14) imply that
\begin{equation}
\mu= 3k_{\rm B}T b^{-2},
\end{equation}
which means that the conventional stiffness $\mu$
of a chain with an arbitrary end-to-end distribution
function is a merely geometrical parameter
that is inversely proportional to the square
of the average end-to-end distance $b$.
According to Eq. (15), the stiffness of a Gaussian
chain with the mean square end-to-end distance $b_{G}$
reads
\begin{equation}
\mu_{G}=3k_{\rm B}T b_{G}^{-2} .
\end{equation}
Obviously, Eq. (16) coincides with Eq. (10) with
the persistence length $l_{\rm p}=\frac{1}{2}b_{0}$.
Setting
$\Psi=\mu/\mu_{G},$
we find from Eqs. (15) and (16) that
\begin{equation}
\Psi=\Bigl (\frac{b_{G}}{b}\Bigr )^{2}.
\end{equation}

\section{Strain energy of a chain in an ensemble}

Our aim now to derive a formula for the strain energy
of a chain in an ensemble of macromolecules
whose deformation at the micro-level coincides with
macro-deformation (the affinity hypothesis).
To simplify the analysis, we assume the deformation
to be incompressible and adopt the conventional
hypothesis that inter-chain interactions may be
accounted for by the incompressibility condition.

Denote by ${\bf Q}$ the end-to-end vector of
a chain in the initial (reference) state and
by ${\bf q}$ the end-to-end vector in the actual
(deformed) state at an arbitrary instant $t\geq 0$.
The distribution functions of end-to-end vectors in
the initial and deformed states read
$p_{0}({\bf Q})$ and $p(t,{\bf q})$, respectively.
Transformation of the reference state of the chain
into its deformed state is described by
\begin{equation}
{\bf q}={\bf F}(t)\cdot {\bf Q},
\end{equation}
where ${\bf F}$ is the deformation gradient for
macro-deformation, and the dot denotes inner product.
The function ${\bf F}(t)$ obeys the differential
equation
\begin{equation}
\frac{d{\bf F}}{dt} ={\bf L}\cdot {\bf F},
\qquad
{\bf F}(0)={\bf I},
\end{equation}
where ${\bf L}(t)$ is the velocity gradient,
and ${\bf I}$ is the unit tensor.
For an incompressible macro-deformation,
the Smoluchowski equation for the function $p(t,{\bf q})$
reads \cite{DE86}
\begin{equation}
\frac{\partial p}{\partial t}
=-\frac{\partial p}{\partial {\bf q}}
\cdot {\bf L}\cdot {\bf q} ,
\qquad
p(0,{\bf q})=p_{0}({\bf q}).
\end{equation}
The solution of Eq. (20) is given by
\begin{equation}
p(t,{\bf q})= p_{0}({\bf F}^{-1}(t)\cdot {\bf q}),
\end{equation}
and it satisfies the normalization condition
\begin{equation}
\int p(t,{\bf q})d{\bf q}=1.
\end{equation}
The distribution functions $p_{0}({\bf Q})$ and
$p(t,{\bf q})$ are expressed in terms of appropriate
configurational free energies $U_{0}({\bf Q})$
and $U(t,{\bf q})$ by the Boltzmann relations
\begin{equation}
p_{0}({\bf Q})=\exp \Bigl [-\frac{U_{0}({\bf Q})}{k_{\rm B}T}
\Bigr ],
\qquad
p(t,{\bf q})=\exp \Bigl [-\frac{U(t,{\bf q})}{k_{\rm B}T}
\Bigr ].
\end{equation}
There are two ways to determine the strain energy of a
chain $W$.
According to the first, we calculate
the increment $\Delta U$ of the configurational
free energy caused by transition from the
reference state to the actual state,
\[
\Delta U(t,{\bf Q})=U(t,{\bf Q})-U_{0}({\bf Q})
=-k_{\rm B}T \Bigl [ \ln p(t,{\bf Q})
-\ln p_{0}({\bf Q})\Bigr ],
\]
and average it with the help of the distribution function
in the reference state,
\begin{equation}
W_{1}(t) = -k_{\rm B}T \int \Bigl [
\ln p(t,{\bf Q})-\ln p_{0}({\bf Q})\Bigr ]
p_{0}({\bf Q}) d{\bf Q}.
\end{equation}
Following the other approach,
the increment of the configurational free energy
is calculated with respect to the actual state,
\[
\Delta U(t,{\bf q})=U_{0}({\bf q})-U(t,{\bf q})
=-k_{\rm B}T \Bigl [ \ln p_{0}({\bf q})-
\ln p(t,{\bf q})\Bigr ],
\]
and it is averaged by using the distribution
function in the deformed state,
\begin{equation}
W_{2}(t) = -k_{\rm B}T \int \Bigl [
\ln p_{0}({\bf q})-\ln p(t, {\bf q})\Bigr ]
p(t,{\bf q}) d{\bf q}.
\end{equation}
Substitution of Eq. (21) into Eq. (24) results in
\begin{equation}
W_{1}(t)= -k_{\rm B}T \int \Bigl [
\ln p_{0}({\bf F}^{-1}(t)\cdot {\bf Q})
-\ln p_{0}({\bf Q})\Bigr ]p_{0}({\bf Q})d{\bf Q}.
\end{equation}
Combining Eqs. (21) and (25), we find that
\[
W_{2}(t) = k_{\rm B}T
\int \Bigl [ \ln p_{0}({\bf F}^{-1}(t)\cdot {\bf q})
-\ln p_{0}({\bf q})\Bigr ]
p_{0}({\bf F}^{-1}(t) \cdot {\bf q}) d{\bf q}.
\]
Introducing the variable ${\bf Q}$ by Eq. (18)
and bearing in mind that $d{\bf q}=d{\bf Q}$ for
an incompressible deformation, we arrive at
the formula
\begin{equation}
W_{2}(t)= k_{\rm B}T \int \Bigl [ \ln p_{0} ({\bf Q})
-\ln p_{0}({\bf F}(t)\cdot {\bf Q})
\Bigr ] p_{0}({\bf Q}) d{\bf Q}.
\end{equation}
It seems natural to define the strain energy of a chain
$W$ as the weighted sum of the strain energies $W_{1}$
and $W_{2}$ calculated by using different ways
of averaging of the configurational free energy,
\[
W=(1-a) W_{1}+a W_{2},
\]
where $a\in [0,1]$ is a material parameter.
Substitution of Eqs. (26) and (27) into this formula
implies that
\begin{equation}
W(t) = k_{\rm B}T
\int \biggl [ a \Bigl ( \ln p_{0} ({\bf Q})
-\ln p_{0}({\bf F}(t)\cdot {\bf Q})
\Bigr )
+(1-a)\Bigl (\ln p_{0}({\bf Q})
-\ln p_{0}({\bf F}^{-1}(t)\cdot {\bf Q})
\Bigr )\biggr ]p_{0}({\bf Q})d{\bf Q}.
\end{equation}
For an isotropic distribution of end-to-end vectors
in the reference state, we set
\begin{equation}
p_{0}({\bf Q})=P(Q^{2}),
\end{equation}
where $Q^{2}={\bf Q}\cdot {\bf Q}$,
and $P(r)$ is a given function of a scalar argument $r$.
Combining Eqs. (28) and (29) and taking into account that
\[
\Bigl ({\bf F}\cdot {\bf Q})\cdot
\Bigl ({\bf F}\cdot {\bf Q}\Bigr )
={\bf Q}\cdot {\bf C}\cdot {\bf Q},
\quad
\Bigl ({\bf F}^{-1}\cdot {\bf Q}\Bigr )
\cdot \Bigl ({\bf F}^{-1}\cdot {\bf Q}\Bigr )
={\bf Q}\cdot {\bf B}^{-1}\cdot {\bf Q},
\]
where the left and right Cauchy--Green
deformation tensors read
\begin{equation}
{\bf B}={\bf F}\cdot {\bf F}^{\top},
\qquad
{\bf C}={\bf F}^{\top} \cdot {\bf F},
\end{equation}
and $\top$ stands for transpose, we find that
\[
W = k_{\rm B}T \int \biggl [ a \Bigl (\ln P (Q^{2})
-\ln P({\bf Q}\cdot {\bf C}\cdot {\bf Q})\Bigr )
+(1-a)\Bigl ( \ln P(Q^{2}-\ln P({\bf Q}\cdot
{\bf B}^{-1}\cdot {\bf Q})
\Bigr )\biggr ] P(Q^{2})d{\bf Q}.
\]
It follows from this equality and the identity
\[
\int \ln P({\bf Q}\cdot {\bf B}^{-1}
\cdot {\bf Q}) P(Q^{2})d{\bf Q}
=\int \ln P({\bf Q}\cdot {\bf C}^{-1}
\cdot {\bf Q}) P(Q^{2})d{\bf Q}
\]
that the strain energy per chain reads
\begin{equation}
W = k_{\rm B}T \int \biggl [ a \Bigl (\ln P (Q^{2})
-\ln P({\bf Q}\cdot {\bf C}\cdot {\bf Q})\Bigr )
+(1-a)\Bigl ( \ln P(Q^{2})-\ln P({\bf Q}\cdot
{\bf C}^{-1}\cdot {\bf Q})
\Bigr )\biggr ] P(Q^{2})d{\bf Q} .
\end{equation}
Neglecting the energy of inter-chain interaction,
we calculate the strain energy per unit volume
of an ensemble of chains $\tilde{W}$ as the sum
of the strain energies of individual chains,
\begin{equation}
\tilde{W} = k_{\rm B}T M \int \biggl [ a \Bigl (\ln P (Q^{2})
-\ln P({\bf Q}\cdot {\bf C}\cdot {\bf Q})\Bigr )
+(1-a)\Bigl ( \ln P(Q^{2})-\ln P({\bf Q}\cdot
{\bf C}^{-1}\cdot {\bf Q})
\Bigr )\biggr ] P(Q^{2})d{\bf Q} ,
\end{equation}
where $M$ is the number of chains per unit volume.

\section{Elastic modulus of a chain}

Our aim now is to apply Eq. (32) in order to determine
the elastic modulus of an ensemble of chains
under uniaxial tension
\begin{equation}
x_{1}=kX_{1},
\qquad
x_{2}=k^{-\frac{1}{2}}X_{2},
\qquad
x_{3}=k^{-\frac{1}{2}}X_{3},
\end{equation}
where $k$ is an elongation ratio,
$\{ X_{i} \}$ are Cartesian coordinates
in the reference state, and
$\{ x_{i} \}$ are Cartesian coordinates
in the deformed state ($i=1,2,3$).
It follows from Eqs. (18), (30) and (33) that
\begin{equation}
{\bf C}=k^{2}{\bf e}_{1}{\bf e}_{1}
+k^{-1}({\bf e}_{2}{\bf e}_{2}+
{\bf e}_{3}{\bf e}_{3}),
\qquad
{\bf C}^{-1}=k^{-2}{\bf e}_{1}{\bf e}_{1}
+k({\bf e}_{2}{\bf e}_{2}+
{\bf e}_{3}{\bf e}_{3}),
\end{equation}
where ${\bf e}_{i}$ are base vectors of the Cartesian
frame $\{ X_{i} \}$.
Substituting expressions (34) into Eq. (32) and
introducing spherical coordinates $\{ Q,\phi,\theta\}$,
whose $z$ axis coincides with ${\bf e}_{1}$,
we find that
\begin{eqnarray*}
\frac{\tilde{W}}{k_{\rm B}T M} &=&
\int_{0}^{\infty} P(Q^{2}) Q^{2} d Q
\int_{0}^{2\pi} d\phi
\int_{0}^{\pi} \biggl [ a \Bigl (\ln P (Q^{2})
-\ln P(Q^{2} (k^{2}\cos^{2}\theta+
k^{-1}\sin^{2}\theta))\Bigr )
\nonumber\\
&&+(1-a)\Bigl ( \ln P(Q^{2})-\ln P(Q^{2}
(k^{-2}\cos^{2}\theta+k\sin^{2}\theta))
\Bigr )\biggr ] \sin\theta d\theta.
\end{eqnarray*}
Performing integration over $\phi$ and introducing
the new variable $x=\cos\theta$, we obtain
\begin{eqnarray}
\frac{\tilde{W}}{k_{\rm B}T M} &=&
4\pi \int_{0}^{\infty} P(Q^{2}) Q^{2} d Q
\int_{0}^{1} \biggl [ a \Bigl (\ln P (Q^{2})
-\ln P(Q^{2} (k^{2}x^{2}+k^{-1}(1-x^{2})))\Bigr )
\nonumber\\
&&+(1-a)\Bigl ( \ln P(Q^{2})-\ln P(Q^{2}
(k^{-2}x^{2}+k(1-x^{2})))\Bigr )\biggr ] dx.
\end{eqnarray}
Equation (35) provides an exact formula for the
strain energy density of an ensemble of chains
under uniaxial tension.
Simple algebra (see Appendix) implies that at
small strains, when
\begin{equation}
k=1+\epsilon
\qquad
(\epsilon\ll 1),
\end{equation}
this equation reads
\begin{equation}
\frac{\tilde{W}}{k_{\rm B}TM}=4\pi \epsilon^{2}
(K_{1}+K_{2}),
\end{equation}
where terms beyond the second order of smallness
are disregarded, and the coefficients $K_{1}$
and $K_{2}$ are given by (the prime denotes the differentiation)
\begin{equation}
K_{1}=-\int_{0}^{\infty} P^{\prime}(Q^{2})Q^{4} dQ,
\qquad
K_{2}=\frac{2}{5} \int_{0}^{\infty}
\biggl [ \frac{(P^{\prime}(Q^{2}))^{2}}{P(Q^{2})}
-P^{\prime\prime}(Q^{2})\biggr ]Q^{6}dQ .
\end{equation}
It is worth noting that the dependence of $\tilde{W}$
on the parameter $a$ disappears at small strains.
It follows from Eq. (29) that for an isotropic Green
function $G$,
\begin{equation}
P(Q^{2})=G_{\ast}(Q),
\qquad
P^{\prime}(Q^{2})=\frac{G_{\ast}^{\prime}(Q)}{2Q},
\qquad
P^{\prime\prime}(Q^{2})=\frac{1}{4Q^{2}}
\Bigl [ G_{\ast}^{\prime\prime}(Q)
-\frac{G_{\ast}^{\prime}(Q)}{Q}\Bigr ].
\end{equation}
Equations (38) and (39) imply that
\[
K_{1} = -\frac{1}{2} \int_{0}^{\infty} G_{\ast}^{\prime}(Q)
Q^{3} dQ,
\quad
K_{2} = \frac{1}{10}
\int_{0}^{\infty} \frac{(G_{\ast}^{\prime}(Q))^{2}}
{G_{\ast}(Q)}Q^{4} dQ
-\frac{1}{10} \int_{0}^{\infty} \Bigl (
G_{\ast}^{\prime\prime}(Q)Q^{4}
-G_{\ast}^{\prime}(Q) Q^{3} \Bigr )dQ .
\]
Bearing in mind that
\[
\int_{0}^{\infty} G_{\ast}^{\prime\prime}(Q)Q^{4} dQ
=-4\int_{0}^{\infty} G_{\ast}^{\prime}(Q) Q^{3} dQ,
\]
we find that
\[
K_{2} = \frac{1}{10}
\int_{0}^{\infty} \frac{(G_{\ast}^{\prime}(Q))^{2}}
{G_{\ast}(Q)}Q^{4} dQ
+\frac{1}{2} \int_{0}^{\infty}
G_{\ast}^{\prime}(Q) Q^{3}dQ .
\]
Substitution of these expressions into Eq. (37)
results in
\begin{equation}
\tilde{W}=\frac{2}{5} \pi k_{\rm B}TM \epsilon^{2}
\int_{0}^{\infty} \frac{(G_{\ast}^{\prime}(Q))^{2}}
{G_{\ast}(Q)}Q^{4} dQ .
\end{equation}
At uniaxial tension of an incompressible medium,
the strain energy per unit volume $\tilde{W}$ reads
\begin{equation}
\tilde{W}=\tilde{E}\epsilon^{2} ,
\end{equation}
where $\tilde{E}$ denotes Young's modulus.
It is worth noting that Eq. (41) does not
contain the coefficient $\frac{1}{2}$ on the
right-hand side that conventionally arises at small
uniaxial deformations.
It follows from Eqs. (40) and (41) that
\begin{equation}
\tilde{E}=\frac{2}{5} \pi k_{\rm B}TM
\int_{0}^{\infty} \frac{(G_{\ast}^{\prime}(Q))^{2}}
{G_{\ast}(Q)}Q^{4} dQ .
\end{equation}
For an ensemble of Gaussian chains with the end-to-end
distribution function
\begin{equation}
G_{\ast}(Q)
=\Bigl (\frac{3}{2\pi b_{G}^{2}}\Bigr )^{\frac{3}{2}}
\exp \Bigl (-\frac{3Q^{2}}{2b_{G}^{2}}\Bigr ),
\end{equation}
Eq. (42) reads
\[
\tilde{E}_{G} = \frac{18}{5b_{G}^{4}} \pi k_{\rm B}TM
\Bigl (\frac{3}{2\pi b_{G}^{2}}\Bigr )^{\frac{3}{2}}
\int_{0}^{\infty}
\exp \Bigl (-\frac{3Q^{2}}{2b_{G}^{2}}\Bigr )Q^{6}dQ .
\]
Setting $z=Q\sqrt{3}/b_{G}$ and calculating the integral,
we find that
\begin{equation}
\tilde{E}_{G} =\frac{3}{2} k_{\rm B}TM.
\end{equation}

\section{Stiffness of a polymer chain}

Our aim now is to determine stiffness of a polymer chain
based on Eq. (42),
which coincides (to some extent) with Eq. (9).
The remark in the parentheses refers to the fact
that definition (9), widely used in the statistical
physics of macromolecules, differs from that
in mechanical engineering where the stiffness is
measured as the ratio of an appropriate force
to a strain (not to a displacement).
To avoid this discrepancy, we replace $\langle Q\rangle$
in Eq. (9) by the ratio $\langle Q\rangle/b$,
which, together with Eq. (16), implies that the
conventional stiffness of a Gaussian chain is given by
\begin{equation}
\mu^{\circ}_{G}=3k_{\rm B}T b_{G}^{-3}.
\end{equation}
To obtain a counterpart of Eq. (45) grounded on
Eq. (42), it seems natural to introduce a hypothetical
ensemble of closely packed chains (the latter means
that the number of chains per unit volume $M$ equals
$v^{-1}$, where $v=\frac{4}{3}\pi b^{3}$ is the average
volume occupied by a chain), and to define the stiffness
of a chain $S$ as the Young's modulus of this ensemble.
If follows from Eq. (44) that for a Gaussian chain,
\begin{equation}
S_{G}=\frac{9 k_{B}T}{8\pi b_{G}^{3}},
\end{equation}
which differs from Eq. (45) by an insignificant
pre-factor (of order of unity) only.
In the general case, this definition implies that
\begin{equation}
S=\frac{3 k_{B}T}{10 b^{3}}
\int_{0}^{\infty} \frac{(G_{\ast}^{\prime}(Q))^{2}}
{G_{\ast}(Q)}Q^{4} dQ .
\end{equation}
Excluding the coefficient $b$ with the help
of Eq. (11), we arrive at
\begin{equation}
S=\frac{3 k_{\rm B}T}{10}
\int_{0}^{\infty} \frac{(G_{\ast}^{\prime}(Q))^{2}}
{G_{\ast}(Q)}Q^{4} dQ
\biggl [ \frac{\int_{0}^{\infty} G_{\ast}(Q) Q^{2} dQ}
{\int_{0}^{\infty} G_{\ast}(Q) Q^{4} dQ}\biggr ]^{\frac{3}{2}}.
\end{equation}
Formula (48) allows the stiffness of a chain with
an arbitrary end-to-end distribution function
$G_{\ast}(Q)$ to be calculated.
Introducing the relative stiffness $\Phi$
as the ratio of the chain stiffness to that for
a Gaussian chain, $\Phi=S/S_{G}$,
we find from Eqs. (17), (46) and (47) that
\begin{equation}
\Phi=\frac{4\pi}{15} \Psi^{\frac{3}{2}}
\int_{0}^{\infty} \frac{(G_{\ast}^{\prime}(Q))^{2}}
{G_{\ast}(Q)}Q^{4} dQ .
\end{equation}

\section{Examples}

Our purpose now is to calculate the ratio $\Phi$
for several distribution functions $G_{\ast}(Q)$
and to show that Eq. (49) leads to physically
plausible dependencies of $\Phi$ on material
parameters.

\subsection{A flexible chain with excluded-volume interactions}

The end-to-end distribution function of
a flexible chain modeled as a self-avoiding random
walk may be approximated by the stretched exponential
function \cite{MM71,Clo74},
\begin{equation}
G_{\ast}(Q)=g \exp \Bigl [
-\Bigl (\frac{Q}{l}\Bigr )^{2\delta}\Bigr ] .
\end{equation}
Here $\delta$ and $l$ are positive constants,
and the pre-factor $g$ is found from
the normalization condition
\[
g=\frac{1}{4\pi} \biggl [ \int_{0}^{\infty}
\exp \Bigl (-\Bigl (\frac{Q}{l}\Bigr )^{2\delta}\Bigr )
Q^{2}dQ\biggr ]^{-1}=\frac{\delta}{2\pi l^{3}
\Gamma(\frac{3}{2\delta})},
\]
where
$\Gamma(x)=\int_{0}^{\infty} \exp(-z) z^{x-1} dz$
is the Euler gamma function.
Substitution of expression (50) into Eq. (48)
implies that
\begin{equation}
S=\frac{9k_{\rm B}T}{40 \pi l^{3}}(3+2\delta)
\biggl [ \frac{\Gamma(\frac{3}{2\delta})}
{\Gamma(\frac{5}{2\delta})} \biggr ]^{\frac{3}{2}},
\end{equation}
When $\delta=1$, we set $l=b_{G}\sqrt{\frac{2}{3}}$
in accord with Eqs. (43) and (50).
In this case, Eq. (51) is reduced
to Eq. (46) for the stiffness of a Gaussian chain.
It follows from Eqs. (46) and (51) that
\begin{equation}
\Phi=\frac{2\delta+3}{5}\biggl [
\frac{3\Gamma(\frac{3}{2\delta})}
{2\Gamma(\frac{5}{2\delta})} \biggr ]^{\frac{3}{2}}.
\end{equation}
Setting $\delta\to 0$ in Eq. (52), we find that
$\Phi\to 0$ due to the decay of the expression
in the square brackets.
Bearing in mind that
\[
\lim_{\delta\to \infty} \frac{\Gamma(\frac{3}{2\delta})}
{\Gamma(\frac{5}{2\delta})}=\frac{5}{3},
\]
we conclude that
\[
\lim_{\delta\to\infty} \frac{\Phi}{\delta}
=\sqrt{\frac{5}{2}},
\]
which means that $\Phi$ increases linearly with
$\delta$ at sufficiently large values of the exponent.
To analyze the effect of $\delta$ on the relative
stiffness of a chain, we calculate $\Phi$ for
$\delta\in [1,2]$, i.e. in the interval where typical
values of this parameter are located.
The results of numerical simulation are presented
in Figure 1, which shows that the stiffness monotonically
increases (practically linearly) with $\delta$.
As $\delta$ may be treated as a measure of strength
of excluded-volume interactions, this implies that
repulsive segment interactions cause the growth of
chain stiffness.
It is worth mentioning that our conclusion contradicts
Eq. (17): according to the latter formula,
an increase in $\delta$ causes the growth of the average
end-to-end distance $b$, which, in turn, induces a
decrease in the conventional dimensionless stiffness $\Psi$.

\subsection{A Gaussian chain with electric charges at the ends}

Our aim now is to assess the effect of an electrostatic
field on the stiffness of a Gaussian chain with two equal
charges $e$ fixed at its ends.
It is assumed that the mean square end-to-end distance
of the chain $b_{G}$ is smaller than the Debye screening
length $l_{D}$, which implies that the Coulomb interaction
between charges is not screened \cite{ZRG03}.
The Hamiltonian of a non-charged Gaussian chain
reads
\[
H_{0}=\frac{3k_{\rm B}T}{2b_{0}}\int_{0}^{L}
\Bigl (\frac{d {{\bf r}}}{ds}(s)\Bigr )^{2} ds.
\]
To account for the energy of interaction between
the charges, we replace $H_{0}$ by the Hamiltonian
\begin{equation}
H=H_{0}+\frac{e^{2}}{\varepsilon |{\bf Q}(L)|},
\end{equation}
where $\varepsilon$ is the dielectric constant of
an ion-free dilute solvent in which the chain is
immersed.
The last term on the right-hand side of Eq. (53)
describes the energy of electrostatic repulsion of charges.
Substituting Eq. (53) into Eq. (1) and bearing in mind that
the last term is independent of the curve ${\bf r}(s)$,
we find that
\begin{equation}
G({\bf Q})= g\exp \Bigl (-\frac{3Q^{2}}{2b_{G}^{2}}\Bigr )
\exp \Bigl (-\frac{l_{B}}{Q}\Bigr ) .
\end{equation}
The first exponent in Eq. (54) stands for the
Green function for a neutral Gaussian chain,
$l_{\rm B}=e^{2}/(\varepsilon k_{\rm B}T)$ is
the Bjerrum length, and the constant $g$ is determined
by Eq. (3),
\[
g=\frac{1}{4\pi} \Bigl (\frac{3}{b_{G}^{2}}\Bigr )^{\frac{3}{2}}
\biggl \{ \int_{0}^{\infty} \exp \Bigl [ -\Bigl (\frac{z^{2}}{2}
+\frac{\xi}{z}\Bigr )\Bigr ] z^{2} dz \biggr \}^{-1}
\]
with $\xi=l_{B}\sqrt{3}/b_{G}$.
Formula (54) correctly predicts that the probability
to find the end $s=L$ of the chain in the close vicinity
of the origin (the position of the other end $s=0$)
strongly decreases due to repulsion of charges.
It follows from Eqs. (46), (48) and (54) that
\begin{equation}
\Phi(\xi)=\frac{\sqrt{3}\int_{0}^{\infty}
\exp(-\frac{z^{2}}{2}-\frac{\xi}{z})(z^{3}-\xi)^{2} dz}
{5\int_{0}^{\infty}\exp(-\frac{z^{2}}{2}-\frac{\xi}{z})z^{2} d z}
\biggl [\frac{\int_{0}^{\infty} \exp(-\frac{z^{2}}{2}
-\frac{\xi}{z})z^{2}d z}{\int_{0}^{\infty}\exp(-\frac{z^{2}}{2}
-\frac{\xi}{z})z^{4} dz} \biggr ]^{\frac{3}{2}}.
\end{equation}
The function $\Phi(\xi)$ is plotted
(in the double logarithmic coordinates with
$\log=\log_{10}$) in Figure 2.
According to this figure, $\Phi$ decreases
with $\xi$.
At sufficiently large $\xi$, the curve $\Phi(\xi)$
may be approximated by the dependence
\begin{equation}
\log \Phi =\Phi_{0}-\Phi_{1}\log \xi ,
\end{equation}
where the coefficients $\Phi_{0}$ and $\Phi_{1}$
are determined by the least-squares method.
Our results of numerical analysis demonstrate
that $\Phi_{1}\approx \frac{1}{3}$, which implies
the scaling law
\begin{equation}
\Phi \propto \Bigl (\frac{b_{G}}{l_{\rm B}}
\Bigr )^{\frac{1}{3}}
\quad
(l_{B}\gg b_{G}).
\end{equation}
The conclusion that the stiffness of a charged Gaussian
chain is smaller than that of an appropriate neutral chain
may be explained by the fact that electrostatic repulsion
of chain ends increases the end-to-end distance in
the reference state, which means that the number of available
configurations, and, as a consequence, the chain entropy
are reduced.
It is worth noting, however, that the same explanation
is inapplicable to flexible chains with excluded-volume
interactions, because Eq. (52) demonstrates the growth
of stiffness due to segment interactions.

\subsection{A flexible chain with the maximum stiffness}

It is of interest to determine the distribution
function of end-to-end vectors for a chain whose
stiffness is maximal.
The analysis is confined to smooth radial
distribution functions $G_{\ast}(Q)$,
which are positive at any $Q\in [0,\infty)$,
tend to zero rather rapidly as $Q\to \infty$,
and remain bounded together with their derivatives
in the vicinity of $Q=0$.
We suppose also that the second moment of the
distribution function $\langle Q^{2} \rangle=b_{\rm c}^{2}$
and the most probable end-to-end distance $l_{\rm c}$
are fixed.
The latter quantity is determined from the condition
of maximum of the function $G_{\ast}(Q)Q^{2}$,
\[
l_{\rm c}=\arg\max_{Q} G_{\ast}(Q)Q^{2}.
\]
To demonstrate that these conditions uniquely
determine the radial distribution function of
a chain with the maximum stiffness,
we present Eq. (47) in the form
\begin{equation}
S=\frac{3k_{\rm B}T}{40 \pi b_{\rm c}^{3}}
{\cal F}(G_{\ast}(Q)),
\end{equation}
where the functional ${\cal F}$ reads
\begin{equation}
{\cal F}(y(Q))=
\int_{0}^{\infty} \frac{(y^{\prime}(Q))^{2}}{y(Q)}Q^{4}dQ
\biggl [\int_{0}^{\infty} y(Q) Q^{2} dQ\biggr ]^{-1}.
\end{equation}
The presence of the last term in Eq. (59) allows
arbitrary (non-normalized) Green functions $y(Q)$ to be
considered.
Denote by $\lambda$ the maximum of the functional
${\cal F}(y)$ on the set of smooth positive functions
$y(Q)$.
A function $y_{0}(Q)$ that maximizes ${\cal F}(y)$
satisfies the equality
\begin{equation}
\int_{0}^{\infty} \biggl [
\frac{(y_{0}^{\prime}(Q))^{2}Q^{4}}{y_{0}(Q)}
-\lambda y_{0}(Q)Q^{2}\biggr ] dQ=0.
\end{equation}
The first term in Eq. (60) is transformed by integration
by parts
\[
\int_{0}^{\infty} \frac{(y_{0}^{\prime}(Q))^{2}Q^{4}}{y_{0}(Q)} dQ
=\int_{0}^{\infty} \frac{y_{0}^{\prime}(Q)Q^{4}}{y_{0}(Q)}
dy_{0}(Q)
=y_{0}^{\prime}(Q)Q^{4}\biggl |_{0}^{\infty}
-\int_{0}^{\infty} y_{0}(Q)\Bigl (
\frac{y_{0}^{\prime}(Q)Q^{4}}{y_{0}(Q)}\Bigr )^{\prime} dQ.
\]
Our assumptions regarding the properties of the
function $y_{0}(Q)$ imply that
the out-of-integral term vanishes.
Combining this equality with Eq. (60), we obtain
\[
\int_{0}^{\infty} \biggl [
\Bigl ( \frac{y_{0}^{\prime}(Q)Q^{4}}{y_{0}(Q)}\Bigr )^{\prime}
+\lambda Q^{2}\biggr ] y_{0}(Q) dQ=0.
\]
This equation is fulfilled provided that the function
$y_{0}(Q)$ obeys the differential equation
\begin{equation}
\Bigl ( \frac{y_{0}^{\prime}(Q)Q^{4}}{y_{0}(Q)}
\Bigr )^{\prime} +\lambda Q^{2}=0.
\end{equation}
Integration of Eq. (61) implies that
\begin{equation}
\frac{y_{0}^{\prime}(Q)}{y_{0}(Q)}
=\frac{B}{Q^{4}}-\frac{\lambda}{3Q},
\end{equation}
where $B$ is an arbitrary constant.
The general solution of Eq.
(62) reads
\begin{equation}
y_{0}(Q)=A Q^{-\frac{\lambda}{3}}
\exp \Bigl (-\frac{B}{3Q^{3}}\Bigr ),
\end{equation}
where $A$ is another constant.
The parameters $A$, $B$ and $\lambda$ are found
from the equalities
\begin{equation}
4\pi \int_{0}^{\infty} y_{0}(Q) Q^{2} dQ=1,
\quad
4\pi \int_{0}^{\infty} y_{0}(Q) Q^{4} dQ=b_{\rm c}^{2},
\quad
\Bigl [y_{0}(Q)Q^{2}\Bigr ]^{\prime}_{Q=l_{\rm c}}=0,
\end{equation}
which describe the normalization condition for the
distribution function,
the definition of the mean square end-to-end distance,
and the definition of the most probable end-to-end
distance, respectively.
Substitution of Eq. (63) into the last formula in Eq. (64)
results in
\begin{equation}
B=\frac{\lambda-6}{3}l_{\rm c}^{3}.
\end{equation}
Dividing the second equality in Eq. (64) by the
first and using Eqs. (63) and (65), we find that
\[
b_{\rm c}^{2}=\frac{\int_{0}^{\infty}
\exp(-\frac{(\lambda-6)l_{\rm c}^{3}}{9Q^{3}})
Q^{-\frac{\lambda-12}{3}}dQ}
{\int_{0}^{\infty}
\exp(-\frac{(\lambda-6)l_{\rm c}^{3}}{9Q^{3}})
Q^{-\frac{\lambda-6}{3}}dQ}.
\]
Setting $z=Q/l_{\rm c}$ and calculating the integrals,
we arrive at the formula
\begin{equation}
\Bigl (\frac{b_{\rm c}}{l_{\rm c}}\Bigr )^{2}
=\Bigl (\frac{\lambda-6}{9}\Bigr )^{\frac{2}{3}}
\frac{\Gamma(\frac{\lambda-15}{9})}
{\Gamma(\frac{\lambda-9}{9})}.
\end{equation}
For a Gaussian chain with
$b_{\rm c}=b_{G}$ and $l_{\rm c}=b_{G}\sqrt{\frac{2}{3}}$,
we set $\lambda=9\lambda_{1}$ and find that $\lambda_{1}$
obeys the transcendental equation
\begin{equation}
\Bigl (\lambda_{1}-\frac{2}{3}\Bigr )^{\frac{2}{3}}
\frac{\Gamma(\lambda_{1}-\frac{5}{3})}{\Gamma(\lambda_{1}-1)}
=\frac{3}{2}.
\end{equation}
It follows from Eqs. (46) and (58) that
\begin{equation}
\frac{S_{G}}{S_{\max}}=\frac{5}{27\lambda_{1}}.
\end{equation}
Solving Eq. (67) for $\lambda_{1}$ numerically
and using Eq. (68), we obtain $\lambda_{1}=3.1809$
and $S_{G}=0.0582\; S_{\max}$.
The result is rather surprising.
It means that Gaussian chains are not so flexible:
the stiffness of a Gaussian chain is about 6\% of
the maximal stiffness of a chain with the same
geometrical parameters.

The difference between the shapes of the radial
distribution functions for a Gaussian chain and
for a chain with the maximum stiffness is
seen in Figure 3, where the results of numerical
simulation are presented for $b_{G}=1.0$.
The shape of a Gaussian chain is described by
Eq. (43), whereas the shape of a chain with
the maximum stiffness is determined by Eq. (63),
where $B$ is given by Eq. (65),
$\lambda$ is calculated from Eq. (67),
and $A$ is found from Eq. (3).
Figure 3 demonstrates that the distribution function
of end-to-end vectors for a chain with the maximum
stiffness vanishes in the vicinity of the point
$Q=0$, has a pronounced maximum at $Q=l_{\rm c}$,
and slowly decreases with $Q$ when $Q>l_{\rm c}$.

It is worth noting the importance of the assumption
regarding the smoothness of the distribution function
$G_{\ast}(Q)$ and its positiveness in $[0,\infty)$.
If these constraints are violated, a distribution function
of end-to-end vectors may be constructed for a chain
with an infinite stiffness.
An example is given by
\begin{equation}
G_{\ast}(Q)=\frac{3}{8\pi l}
\Bigl [1-\Bigl (\frac{Q}{l}\Bigr )^{2}\Bigr ]
\quad
(Q\leq l),
\qquad
G_{\ast}(Q)=0
\quad
(Q>l).
\end{equation}
Formula (69) and similar expressions for the radial
distribution function naturally arise for polymer chains
treated as random walks governed by equations of anomalous
diffusion \cite{MMP01,PMM02}.
It should be emphasized, however, that Eq. (47) is
inapplicable to chains, whose distribution functions
have finite supports.
This is explained by the fact that Eq. (47) is grounded
on the hypothesis regarding an affine deformation of a network,
while the probability $p(t,{\bf q})$ is not defined
by Eq. (20), when the vector ${\bf q}={\bf F}\cdot {\bf Q}$
is located in the domain where the distribution function
$p_{0}({\bf Q})$ vanishes.

\subsection{A semi-flexible chain}

Semi-flexible chains provide another class of
macromolecules, for which the applicability of
Eqs. (47) and (48) may be questioned.
This is explained by the fact that the integral
in Eq. (32) for the strain energy density diverges.
On the other hand, appropriate integrals in Eq. (48)
converge, and it is tempting to employ
this formula in order to calculate the stiffness.
Our aim now is to assess the effect of
persistence length $l_{\rm p}$ on the stiffness
of worm-like chains and to demonstrate that Eq. (48)
results in a physically plausible behavior of
the function $S(l_{\rm p})$.

Assuming that $G_{\ast}(Q)=\Lambda(Q/L)$, where $L$ stands
for the chain length, and setting $z=Q/L$,
we find from Eq. (48) that
\begin{equation}
S=\frac{3 k_{\rm B}T}{10}
\int_{0}^{\infty} \frac{(\Lambda(z))^{2}}{\Lambda(z)}
z^{4} dz
\biggl [ \frac{\int_{0}^{\infty} \Lambda(z) z^{2} dz}
{\int_{0}^{\infty} \Lambda(z) z^{4} dz}\biggr ]^{\frac{3}{2}}.
\end{equation}
We begin with the analysis of semi-flexible chains with
the distribution function \cite{WF96}
\begin{eqnarray}
\Lambda(z) &=& \frac{g}{[\zeta(1-z)]^{\frac{3}{2}}}
\sum_{m=1}^{\infty}
\exp\Bigl [-\Bigl ( \frac{m-\frac{1}{2}}{
(\zeta(1-z))^{\frac{1}{2}}}\Bigr )^{2}\Bigr ]
H_{2}\Bigl (\frac{m-\frac{1}{2}}{(\zeta(1-z))^{\frac{1}{2}}}
\Bigr )
\quad
(z<1),
\nonumber\\
\Lambda(z) &=& 0
\quad
(z\geq 1).
\end{eqnarray}
Here $\zeta=l_{\rm p}/L$ stands for the dimensionless
persistence length, $H_{2}(z)=4z^{2}-2$,
and the pre-factor $g$ is determined by Eq. (3).
We substitute expression (71) into Eq. (70),
calculate the integrals numerically (by the Simpson
method with the step $\Delta z=0.001$)
taking into account 500 terms in the series,
and calculate $S$.
The ratio $\overline{S}=S/(k_{\rm B}T)$ is plotted
versus $\zeta$ in Figure 4 at relatively small
values of $\zeta$ (in the linear scale)
and in Figure 5 at arbitrary $\zeta$ (in the double
logarithmic scale).

The results of numerical simulation based on
Eq. (71) are compared with those found
by using the radial distribution functions proposed
in \cite{HT97,Win03},
\begin{eqnarray}
\Lambda(z) &=& \frac{g}{(1-z^{2})^{\frac{9}{2}}}
\exp \Bigl [-\frac{9}{8\zeta (1-z^{2})}\Bigr ]
\quad
(z<1),
\\
\Lambda(z) &=& \frac{g}{(1-z^{2})^{\frac{3}{2}}(2-z^{2})^{3}}
\exp \Bigl [-\frac{3}{4\zeta (1-z^{2})}\Bigr ]
\quad
(z<1).
\end{eqnarray}
Equations (72) and (73) presume the function $\Lambda(z)$
to vanish at $z\geq 1$, in accord with the second equality
in Eq. (71).
The ratios $\overline{S}$ calculated from Eqs. (70),
(72) and (73) are plotted versus the
dimensionless persistence length $\zeta$ in Figure 5.
The results of numerical analysis are approximated
by the function
\begin{equation}
\log \overline{S}=S_{0}+S_{1}\log \zeta,
\end{equation}
where the coefficients $S_{0}$ and $S_{1}$
are determined by the least-squares technique.
Figure 5 shows that Eq. (74) correctly fits the
dependence $S(\zeta)$ at relatively large values
of $\zeta$.
The coefficient $S_{1}$ in Eq. (74) is close to two
for all models under investigation, which implies
that the stiffness of a worm-like chain grows with
persistence length $l_{\rm p}$ being proportional
to $l_{\rm p}^{2}$ (in other words, being proportional
to the square of the bending stiffness $\kappa$),
as it is expected \cite{MKJ95}.
However, some discrepancies between the scaling
prediction
\begin{equation}
S\propto l_{\rm p}^{\alpha}
\end{equation}
with $\alpha=2$ and the numerical results should be
mentioned.
The exponent $\alpha$ practically equals 2 for
distribution function (72), and it is close
to $\frac{9}{5}$ for functions (71) and (73)
(no changes in $\alpha$ are observed with the
growth of the number of terms in the series
and a decrease in the step $\Delta z$ employed
in numerical integration).
The deviations of the exponent for the
distribution functions (71) and (73)
from $\alpha=2$ may be explained by the fact that
inadequate approximations of the Green function
were chosen in the derivation of these relations.
The method suggested in \cite{WF96} is based on
the hypothesis that any curve ${\bf r}(s)$ in Eq.
(1) is close enough to the straight line connecting
the points ${\bf r}(0)={\bf 0}$ and ${\bf r}(L)={\bf Q}$.
This assumption is correct for end-to-end vectors
${\bf Q}$ with $Q=L$, but it may cause large discrepancies
at $Q<L$ driven by the constraint on the curve length.
Figure 5 reveals that the method of softening
this constraint proposed in \cite{HT97}
(the local restriction on the length of the
tangent vector is replaced by the global one)
leads to a substantially better approximation,
whereas the approach developed in \cite{Win03} does not
improve substantially the quality of approximation.

Based on another way of thinking, the so-called
Dirac chains were introduced in \cite{KV94}
with the radial distribution function
\begin{equation}
\Lambda(z)=\frac{g}{\sqrt{1-z^{2}}}
I_{1}\Bigl (\frac{3\sqrt{1-z^{2}}}{2\zeta} \Bigr )
\quad
(z<1),
\qquad
\Lambda(z) = 0
\quad
(z\geq 1).
\end{equation}
Here $\zeta$ is the dimensionless persistence length,
\[
I_{1}(z)= \frac{2}{z} \sum_{m=1}^{\infty}
\frac{m(\frac{1}{4} z^{2})^{m}}{(m!)^{2}}
\]
is the modified Bessel function,
and the pre-factor $g$ is determined by Eq. (3).
To assess the stiffness of a chain with the distribution
function (76), we substitute this expression
into Eq. (70), calculate the stiffness $S$,
and plot the ratio $\overline{S}$ versus $\zeta$
in Figure 6 together with its approximation by Eq. (74).
This figure shows that at relatively small persistence
lengths, $l_{\rm p}<0.3 L$, the stiffness $S$ increases
with $l_{\rm p}$, but the scaling exponent
$\alpha\approx 1.5$
is lower than that for Eqs. (71) to (73).
With a further increase in the persistence length
$l_{\rm p}$, the stiffness diminishes,
which indicates that the model becomes inappropriate
at relatively large values of $l_{\rm p}$.

\section{Concluding remarks}

A ``mechanical" definition of stiffness is introduced
for a polymer chain, and an explicit formula is
derived to express the stiffness in terms of the radial
distribution function.
According to our approach, a polymer chain is ``embedded"
into an affine ensemble, and the stiffness is defined
as an elastic modulus per chain of the ensemble.

This relation is applied to calculate the
stiffness of (i) a Gaussian chain,
(ii) a flexible chain modeled as a self-avoiding
random walk,
and (iii) a Gaussian chain carrying electric charges
at its ends.
The influence of the Bjerrum length $l_{\rm B}$
on the stiffness $S$ of a charged Gaussian chain
has been studied numerically.
It is revealed that $S$ decreases with $l_{\rm B}$
being proportional to $l_{\rm B}^{-\frac{1}{3}}$.

An analytical formula is derived for the distribution
function of a chain with the maximum stiffness.
It is found that the stiffness of a Gaussian
chain is about 6\% of that for a chain with the maximum
stiffness and the same geometrical parameters.

The effect of persistence length $l_{\rm p}$
on the stiffness of semi-flexible chains
has been evaluated numerically.
It is demonstrated that the stiffness
of worm-like chains increases with $l_{\rm p}$
following the pattern $E\propto l_{\rm p}^{\alpha}$
with $\alpha \approx 2$.
For a Dirac chain, stiffness increases with
persistence length when $l_{\rm p}$ is less
than 30\% of the contour length $L$, and the
scaling exponent $\alpha$ is close to $\frac{3}{2}$.

\section*{Appendix}
\renewcommand{\theequation}{A-\arabic{equation}}
\setcounter{equation}{0}

Neglecting terms beyond the second order of smallness
with respect to $\epsilon$, we find from Eq. (36) that
\begin{eqnarray}
&& k^{2}x^{2}+k^{-1}(1-x^{2})=1-\epsilon (1-3x^{2})
+\epsilon^{2},
\nonumber\\
&& k^{-2}x^{2}+k(1-x^{2})=1+\epsilon(1-3x^{2})+
3\epsilon^{2}x^{2}.
\end{eqnarray}
The expansion of the function $\ln P(Q^{2}(1+\alpha))$
in a Taylor series with respect to $\alpha$
in the vicinity of the point $\alpha=0$ reads
\begin{equation}
\ln P(Q^{2}(1+\alpha))=\ln P(Q^{2})
+\alpha Q^{2}\frac{P^{\prime}(Q^{2})}{P(Q^{2})}
+\frac{1}{2}\alpha^{2}Q^{4}
\frac{P^{\prime\prime}(Q^{2})P(Q^{2})-(P^{\prime}(Q^{2}))^{2}}
{P^{2}(Q^{2})}+\ldots,
\end{equation}
where the prime stands for the derivative of the
function $P$.
It follows from Eqs. (A-1) and (A-2) that
\begin{eqnarray*}
\ln P (Q^{2})&-& \ln P(Q^{2} (k^{2}x^{2}+k^{-1}(1-x^{2})))
=\epsilon Q^{2}(1-3x^{2})\frac{P^{\prime}(Q^{2})}{P(Q^{2})}
-\epsilon^{2}Q^{2}\frac{P^{\prime}(Q^{2})}{P(Q^{2})}
\nonumber\\
&& +\frac{1}{2}\epsilon^{2}Q^{4}(1-3x^{2})^{2}
\frac{(P^{\prime}(Q^{2}))^{2}-P^{\prime\prime}(Q^{2})P(Q^{2})}
{P^{2}(Q^{2})},
\nonumber\\
\ln P (Q^{2}) &-& \ln P(Q^{2} (k^{-2}x^{2}+k(1-x^{2})))
=-\epsilon Q^{2}(1-3x^{2})\frac{P^{\prime}(Q^{2})}{P(Q^{2})}
-3\epsilon^{2}Q^{2}x^{2}\frac{P^{\prime}(Q^{2})}{P(Q^{2})}
\nonumber\\
&& +\frac{1}{2}\epsilon^{2}Q^{4}(1-3x^{2})^{2}
\frac{(P^{\prime}(Q^{2}))^{2}-P^{\prime\prime}(Q^{2})P(Q^{2})}
{P^{2}(Q^{2})}.
\end{eqnarray*}
Integrating these expressions with respect to $x$
and taking into account that
\[
\int_{0}^{1} (1-3x^{2})dx=0,
\qquad
\int_{0}^{1} 3x^{2}dx=1,
\qquad
\int_{0}^{1} (1-3x^{2})^{2} dx=\frac{4}{5},
\]
we find that
\begin{eqnarray}
&& \int_{0}^{1} \Bigl [
\ln P (Q^{2})-\ln P(Q^{2} (k^{2}x^{2}+k^{-1}(1-x^{2})))
\Bigr ] dx
\nonumber\\
&& =\int_{0}^{1} \Bigl [
\ln P (Q^{2})-\ln P(Q^{2} (k^{-2}x^{2}+k(1-x^{2})))
\Bigr ] dx
\nonumber\\
&&= \epsilon^{2}\biggl [
-\frac{P^{\prime}(Q^{2})}{P(Q^{2})}Q^{2}
+\frac{2}{5}\frac{(P^{\prime}(Q^{2}))^{2}
-P^{\prime\prime}(Q^{2})P(Q^{2})}{P^{2}(Q^{2})}
Q^{4}\biggr ].
\end{eqnarray}
Equations (37) and (38) follow from Eqs. (35) and (A-3).

\newpage
\section*{List of figures}
\parindent 0 mm

{\bf Figure 1:}
The dimensionless stiffness $\Phi$ of a flexible
chain with excluded-volume interactions versus the
dimensionless parameter $\delta$.
\vspace*{2 mm}

{\bf Figure 2:}
The dimensionless stiffness $\Phi$ of a Gaussian
chain with electric changes at the ends
versus the dimensionless Bjerrum length $\xi$.
Circles: results of numerical simulation.
Solid line: their approximation by Eq. (56)
with $\Phi_{0}=0.031$ and $\Phi_{1}=0.335$.
\vspace*{2 mm}

{\bf Figure 3:}
The radial distribution functions $G_{\ast}(z)$
with $z=Q/b_{\rm c}$ for a Gaussian chain (solid line)
and for a flexible chain with the maximal
stiffness and the same parameters $b_{\rm c}$
and $l_{\rm c}$ (circles).
\vspace*{2 mm}

{\bf Figure 4:}
The ratio $\overline{S}$ versus the dimensionless
persistence length $\zeta$ for approximation (71)
of the Green function.
\vspace*{2 mm}

{\bf Figure 5:}
The ratio $\overline{S}$ versus the dimensionless
persistence length $\zeta$ for three approximations
of the distribution function.
Symbols: results of numerical simulation.
Unfilled circles: Eq. (71).
Filed circles: Eq. (72).
Diamonds: Eq. (73).
Solid lines: their approximations by Eq. (74).
Curve 1: $S_{0}=1.27$, $S_{1}=1.83$.
Curve 2: $S_{0}=0.91$, $S_{1}=1.99$.
Curve 3: $S_{0}=0.26$, $S_{1}=1.79$.
\vspace*{2 mm}

{\bf Figure 6:}
The ratio $\overline{S}$ versus the dimensionless
persistence length $\zeta$ for a Dirac chain.
Circles: results of numerical simulation.
Solid lines: their approximations by Eq. (74).
Curve 1: $S_{0}=-0.04$, $S_{1}=1.47$.
Curve 2: $S_{0}=-2.87$, $S_{1}=-3.97$.
\vspace*{80 mm}

\setlength{\unitlength}{0.75 mm}
\begin{figure}[tbh]
\begin{center}
\begin{picture}(100,100)
\put(0,0){\framebox(100,100)}
\multiput(10,0)(10,0){9}{\line(0,1){2}}
\multiput(0,20)(0,20){4}{\line(1,0){2}}
\put(0,-9){1.0}
\put(94,-9){2.0}
\put(50,-9){$\delta$}
\put(-10,0){0.0}
\put(-10,96){5.0}
\put(-10,70){$\Phi$}

\put(   0.33,   20.20){\circle*{0.8}}
\put(   0.67,   20.39){\circle*{0.8}}
\put(   1.00,   20.59){\circle*{0.8}}
\put(   1.33,   20.79){\circle*{0.8}}
\put(   1.67,   20.99){\circle*{0.8}}
\put(   2.00,   21.19){\circle*{0.8}}
\put(   2.33,   21.39){\circle*{0.8}}
\put(   2.67,   21.59){\circle*{0.8}}
\put(   3.00,   21.79){\circle*{0.8}}
\put(   3.33,   21.99){\circle*{0.8}}
\put(   3.67,   22.19){\circle*{0.8}}
\put(   4.00,   22.39){\circle*{0.8}}
\put(   4.33,   22.60){\circle*{0.8}}
\put(   4.67,   22.80){\circle*{0.8}}
\put(   5.00,   23.00){\circle*{0.8}}
\put(   5.33,   23.20){\circle*{0.8}}
\put(   5.67,   23.41){\circle*{0.8}}
\put(   6.00,   23.61){\circle*{0.8}}
\put(   6.33,   23.81){\circle*{0.8}}
\put(   6.67,   24.02){\circle*{0.8}}
\put(   7.00,   24.22){\circle*{0.8}}
\put(   7.33,   24.43){\circle*{0.8}}
\put(   7.67,   24.63){\circle*{0.8}}
\put(   8.00,   24.84){\circle*{0.8}}
\put(   8.33,   25.04){\circle*{0.8}}
\put(   8.67,   25.25){\circle*{0.8}}
\put(   9.00,   25.46){\circle*{0.8}}
\put(   9.33,   25.66){\circle*{0.8}}
\put(   9.67,   25.87){\circle*{0.8}}
\put(  10.00,   26.08){\circle*{0.8}}
\put(  10.33,   26.28){\circle*{0.8}}
\put(  10.67,   26.49){\circle*{0.8}}
\put(  11.00,   26.70){\circle*{0.8}}
\put(  11.33,   26.91){\circle*{0.8}}
\put(  11.67,   27.11){\circle*{0.8}}
\put(  12.00,   27.32){\circle*{0.8}}
\put(  12.33,   27.53){\circle*{0.8}}
\put(  12.67,   27.74){\circle*{0.8}}
\put(  13.00,   27.95){\circle*{0.8}}
\put(  13.33,   28.16){\circle*{0.8}}
\put(  13.67,   28.37){\circle*{0.8}}
\put(  14.00,   28.58){\circle*{0.8}}
\put(  14.33,   28.79){\circle*{0.8}}
\put(  14.67,   29.00){\circle*{0.8}}
\put(  15.00,   29.21){\circle*{0.8}}
\put(  15.33,   29.42){\circle*{0.8}}
\put(  15.67,   29.63){\circle*{0.8}}
\put(  16.00,   29.84){\circle*{0.8}}
\put(  16.33,   30.05){\circle*{0.8}}
\put(  16.67,   30.26){\circle*{0.8}}
\put(  17.00,   30.47){\circle*{0.8}}
\put(  17.33,   30.68){\circle*{0.8}}
\put(  17.67,   30.89){\circle*{0.8}}
\put(  18.00,   31.10){\circle*{0.8}}
\put(  18.33,   31.31){\circle*{0.8}}
\put(  18.67,   31.52){\circle*{0.8}}
\put(  19.00,   31.74){\circle*{0.8}}
\put(  19.33,   31.95){\circle*{0.8}}
\put(  19.67,   32.16){\circle*{0.8}}
\put(  20.00,   32.37){\circle*{0.8}}
\put(  20.33,   32.58){\circle*{0.8}}
\put(  20.67,   32.79){\circle*{0.8}}
\put(  21.00,   33.01){\circle*{0.8}}
\put(  21.33,   33.22){\circle*{0.8}}
\put(  21.67,   33.43){\circle*{0.8}}
\put(  22.00,   33.64){\circle*{0.8}}
\put(  22.33,   33.85){\circle*{0.8}}
\put(  22.67,   34.07){\circle*{0.8}}
\put(  23.00,   34.28){\circle*{0.8}}
\put(  23.33,   34.49){\circle*{0.8}}
\put(  23.67,   34.70){\circle*{0.8}}
\put(  24.00,   34.92){\circle*{0.8}}
\put(  24.33,   35.13){\circle*{0.8}}
\put(  24.67,   35.34){\circle*{0.8}}
\put(  25.00,   35.55){\circle*{0.8}}
\put(  25.33,   35.77){\circle*{0.8}}
\put(  25.67,   35.98){\circle*{0.8}}
\put(  26.00,   36.19){\circle*{0.8}}
\put(  26.33,   36.40){\circle*{0.8}}
\put(  26.67,   36.62){\circle*{0.8}}
\put(  27.00,   36.83){\circle*{0.8}}
\put(  27.33,   37.04){\circle*{0.8}}
\put(  27.67,   37.25){\circle*{0.8}}
\put(  28.00,   37.47){\circle*{0.8}}
\put(  28.33,   37.68){\circle*{0.8}}
\put(  28.67,   37.89){\circle*{0.8}}
\put(  29.00,   38.10){\circle*{0.8}}
\put(  29.33,   38.32){\circle*{0.8}}
\put(  29.67,   38.53){\circle*{0.8}}
\put(  30.00,   38.74){\circle*{0.8}}
\put(  30.33,   38.95){\circle*{0.8}}
\put(  30.67,   39.17){\circle*{0.8}}
\put(  31.00,   39.38){\circle*{0.8}}
\put(  31.33,   39.59){\circle*{0.8}}
\put(  31.67,   39.80){\circle*{0.8}}
\put(  32.00,   40.02){\circle*{0.8}}
\put(  32.33,   40.23){\circle*{0.8}}
\put(  32.67,   40.44){\circle*{0.8}}
\put(  33.00,   40.65){\circle*{0.8}}
\put(  33.33,   40.87){\circle*{0.8}}
\put(  33.67,   41.08){\circle*{0.8}}
\put(  34.00,   41.29){\circle*{0.8}}
\put(  34.33,   41.50){\circle*{0.8}}
\put(  34.67,   41.71){\circle*{0.8}}
\put(  35.00,   41.93){\circle*{0.8}}
\put(  35.33,   42.14){\circle*{0.8}}
\put(  35.67,   42.35){\circle*{0.8}}
\put(  36.00,   42.56){\circle*{0.8}}
\put(  36.33,   42.77){\circle*{0.8}}
\put(  36.67,   42.98){\circle*{0.8}}
\put(  37.00,   43.20){\circle*{0.8}}
\put(  37.33,   43.41){\circle*{0.8}}
\put(  37.67,   43.62){\circle*{0.8}}
\put(  38.00,   43.83){\circle*{0.8}}
\put(  38.33,   44.04){\circle*{0.8}}
\put(  38.67,   44.25){\circle*{0.8}}
\put(  39.00,   44.46){\circle*{0.8}}
\put(  39.33,   44.67){\circle*{0.8}}
\put(  39.67,   44.89){\circle*{0.8}}
\put(  40.00,   45.10){\circle*{0.8}}
\put(  40.33,   45.31){\circle*{0.8}}
\put(  40.67,   45.52){\circle*{0.8}}
\put(  41.00,   45.73){\circle*{0.8}}
\put(  41.33,   45.94){\circle*{0.8}}
\put(  41.67,   46.15){\circle*{0.8}}
\put(  42.00,   46.36){\circle*{0.8}}
\put(  42.33,   46.57){\circle*{0.8}}
\put(  42.67,   46.78){\circle*{0.8}}
\put(  43.00,   46.99){\circle*{0.8}}
\put(  43.33,   47.20){\circle*{0.8}}
\put(  43.67,   47.41){\circle*{0.8}}
\put(  44.00,   47.62){\circle*{0.8}}
\put(  44.33,   47.83){\circle*{0.8}}
\put(  44.67,   48.04){\circle*{0.8}}
\put(  45.00,   48.24){\circle*{0.8}}
\put(  45.33,   48.45){\circle*{0.8}}
\put(  45.67,   48.66){\circle*{0.8}}
\put(  46.00,   48.87){\circle*{0.8}}
\put(  46.33,   49.08){\circle*{0.8}}
\put(  46.67,   49.29){\circle*{0.8}}
\put(  47.00,   49.50){\circle*{0.8}}
\put(  47.33,   49.70){\circle*{0.8}}
\put(  47.67,   49.91){\circle*{0.8}}
\put(  48.00,   50.12){\circle*{0.8}}
\put(  48.33,   50.33){\circle*{0.8}}
\put(  48.67,   50.54){\circle*{0.8}}
\put(  49.00,   50.74){\circle*{0.8}}
\put(  49.33,   50.95){\circle*{0.8}}
\put(  49.67,   51.16){\circle*{0.8}}
\put(  50.00,   51.37){\circle*{0.8}}
\put(  50.33,   51.57){\circle*{0.8}}
\put(  50.67,   51.78){\circle*{0.8}}
\put(  51.00,   51.99){\circle*{0.8}}
\put(  51.33,   52.19){\circle*{0.8}}
\put(  51.67,   52.40){\circle*{0.8}}
\put(  52.00,   52.61){\circle*{0.8}}
\put(  52.33,   52.81){\circle*{0.8}}
\put(  52.67,   53.02){\circle*{0.8}}
\put(  53.00,   53.22){\circle*{0.8}}
\put(  53.33,   53.43){\circle*{0.8}}
\put(  53.67,   53.64){\circle*{0.8}}
\put(  54.00,   53.84){\circle*{0.8}}
\put(  54.33,   54.05){\circle*{0.8}}
\put(  54.67,   54.25){\circle*{0.8}}
\put(  55.00,   54.46){\circle*{0.8}}
\put(  55.33,   54.66){\circle*{0.8}}
\put(  55.67,   54.86){\circle*{0.8}}
\put(  56.00,   55.07){\circle*{0.8}}
\put(  56.33,   55.27){\circle*{0.8}}
\put(  56.67,   55.48){\circle*{0.8}}
\put(  57.00,   55.68){\circle*{0.8}}
\put(  57.33,   55.89){\circle*{0.8}}
\put(  57.67,   56.09){\circle*{0.8}}
\put(  58.00,   56.29){\circle*{0.8}}
\put(  58.33,   56.50){\circle*{0.8}}
\put(  58.67,   56.70){\circle*{0.8}}
\put(  59.00,   56.90){\circle*{0.8}}
\put(  59.33,   57.10){\circle*{0.8}}
\put(  59.67,   57.31){\circle*{0.8}}
\put(  60.00,   57.51){\circle*{0.8}}
\put(  60.33,   57.71){\circle*{0.8}}
\put(  60.67,   57.91){\circle*{0.8}}
\put(  61.00,   58.12){\circle*{0.8}}
\put(  61.33,   58.32){\circle*{0.8}}
\put(  61.67,   58.52){\circle*{0.8}}
\put(  62.00,   58.72){\circle*{0.8}}
\put(  62.33,   58.92){\circle*{0.8}}
\put(  62.67,   59.12){\circle*{0.8}}
\put(  63.00,   59.32){\circle*{0.8}}
\put(  63.33,   59.52){\circle*{0.8}}
\put(  63.67,   59.72){\circle*{0.8}}
\put(  64.00,   59.92){\circle*{0.8}}
\put(  64.33,   60.13){\circle*{0.8}}
\put(  64.67,   60.33){\circle*{0.8}}
\put(  65.00,   60.52){\circle*{0.8}}
\put(  65.33,   60.72){\circle*{0.8}}
\put(  65.67,   60.92){\circle*{0.8}}
\put(  66.00,   61.12){\circle*{0.8}}
\put(  66.33,   61.32){\circle*{0.8}}
\put(  66.67,   61.52){\circle*{0.8}}
\put(  67.00,   61.72){\circle*{0.8}}
\put(  67.33,   61.92){\circle*{0.8}}
\put(  67.67,   62.12){\circle*{0.8}}
\put(  68.00,   62.31){\circle*{0.8}}
\put(  68.33,   62.51){\circle*{0.8}}
\put(  68.67,   62.71){\circle*{0.8}}
\put(  69.00,   62.91){\circle*{0.8}}
\put(  69.33,   63.11){\circle*{0.8}}
\put(  69.67,   63.30){\circle*{0.8}}
\put(  70.00,   63.50){\circle*{0.8}}
\put(  70.33,   63.70){\circle*{0.8}}
\put(  70.67,   63.89){\circle*{0.8}}
\put(  71.00,   64.09){\circle*{0.8}}
\put(  71.33,   64.29){\circle*{0.8}}
\put(  71.67,   64.48){\circle*{0.8}}
\put(  72.00,   64.68){\circle*{0.8}}
\put(  72.33,   64.87){\circle*{0.8}}
\put(  72.67,   65.07){\circle*{0.8}}
\put(  73.00,   65.26){\circle*{0.8}}
\put(  73.33,   65.46){\circle*{0.8}}
\put(  73.67,   65.65){\circle*{0.8}}
\put(  74.00,   65.85){\circle*{0.8}}
\put(  74.33,   66.04){\circle*{0.8}}
\put(  74.67,   66.24){\circle*{0.8}}
\put(  75.00,   66.43){\circle*{0.8}}
\put(  75.33,   66.63){\circle*{0.8}}
\put(  75.67,   66.82){\circle*{0.8}}
\put(  76.00,   67.01){\circle*{0.8}}
\put(  76.33,   67.21){\circle*{0.8}}
\put(  76.67,   67.40){\circle*{0.8}}
\put(  77.00,   67.59){\circle*{0.8}}
\put(  77.33,   67.78){\circle*{0.8}}
\put(  77.67,   67.98){\circle*{0.8}}
\put(  78.00,   68.17){\circle*{0.8}}
\put(  78.33,   68.36){\circle*{0.8}}
\put(  78.67,   68.55){\circle*{0.8}}
\put(  79.00,   68.75){\circle*{0.8}}
\put(  79.33,   68.94){\circle*{0.8}}
\put(  79.67,   69.13){\circle*{0.8}}
\put(  80.00,   69.32){\circle*{0.8}}
\put(  80.33,   69.51){\circle*{0.8}}
\put(  80.67,   69.70){\circle*{0.8}}
\put(  81.00,   69.89){\circle*{0.8}}
\put(  81.33,   70.08){\circle*{0.8}}
\put(  81.67,   70.27){\circle*{0.8}}
\put(  82.00,   70.46){\circle*{0.8}}
\put(  82.33,   70.65){\circle*{0.8}}
\put(  82.67,   70.84){\circle*{0.8}}
\put(  83.00,   71.03){\circle*{0.8}}
\put(  83.33,   71.22){\circle*{0.8}}
\put(  83.67,   71.41){\circle*{0.8}}
\put(  84.00,   71.60){\circle*{0.8}}
\put(  84.33,   71.79){\circle*{0.8}}
\put(  84.67,   71.97){\circle*{0.8}}
\put(  85.00,   72.16){\circle*{0.8}}
\put(  85.33,   72.35){\circle*{0.8}}
\put(  85.67,   72.54){\circle*{0.8}}
\put(  86.00,   72.73){\circle*{0.8}}
\put(  86.33,   72.91){\circle*{0.8}}
\put(  86.67,   73.10){\circle*{0.8}}
\put(  87.00,   73.29){\circle*{0.8}}
\put(  87.33,   73.47){\circle*{0.8}}
\put(  87.67,   73.66){\circle*{0.8}}
\put(  88.00,   73.85){\circle*{0.8}}
\put(  88.33,   74.03){\circle*{0.8}}
\put(  88.67,   74.22){\circle*{0.8}}
\put(  89.00,   74.40){\circle*{0.8}}
\put(  89.33,   74.59){\circle*{0.8}}
\put(  89.67,   74.77){\circle*{0.8}}
\put(  90.00,   74.96){\circle*{0.8}}
\put(  90.33,   75.14){\circle*{0.8}}
\put(  90.67,   75.33){\circle*{0.8}}
\put(  91.00,   75.51){\circle*{0.8}}
\put(  91.33,   75.70){\circle*{0.8}}
\put(  91.67,   75.88){\circle*{0.8}}
\put(  92.00,   76.07){\circle*{0.8}}
\put(  92.33,   76.25){\circle*{0.8}}
\put(  92.67,   76.43){\circle*{0.8}}
\put(  93.00,   76.62){\circle*{0.8}}
\put(  93.33,   76.80){\circle*{0.8}}
\put(  93.67,   76.98){\circle*{0.8}}
\put(  94.00,   77.16){\circle*{0.8}}
\put(  94.33,   77.35){\circle*{0.8}}
\put(  94.67,   77.53){\circle*{0.8}}
\put(  95.00,   77.71){\circle*{0.8}}
\put(  95.33,   77.89){\circle*{0.8}}
\put(  95.67,   78.07){\circle*{0.8}}
\put(  96.00,   78.26){\circle*{0.8}}
\put(  96.33,   78.44){\circle*{0.8}}
\put(  96.67,   78.62){\circle*{0.8}}
\put(  97.00,   78.80){\circle*{0.8}}
\put(  97.33,   78.98){\circle*{0.8}}
\put(  97.67,   79.16){\circle*{0.8}}
\put(  98.00,   79.34){\circle*{0.8}}
\put(  98.33,   79.52){\circle*{0.8}}
\put(  98.67,   79.70){\circle*{0.8}}
\put(  99.00,   79.88){\circle*{0.8}}
\put(  99.33,   80.06){\circle*{0.8}}
\put(  99.67,   80.24){\circle*{0.8}}
\put( 100.00,   80.42){\circle*{0.8}}

\end{picture}
\end{center}
\vspace*{10 mm}

\caption{}
\end{figure}

\setlength{\unitlength}{0.75 mm}
\begin{figure}[tbh]
\begin{center}
\begin{picture}(100,100)
\put(0,0){\framebox(100,100)}
\multiput(12.5,0)(12.5,0){7}{\line(0,1){2}}
\multiput(0,10)(0,10){9}{\line(1,0){2}}
\put(0,-9){$-1.0$}
\put(94,-9){3.0}
\put(50,-9){$\log \xi$}
\put(-14,0){$-1.0$}
\put(-10,96){0.0}
\put(-14,70){$\log \Phi$}

\put(   3.33,   98.68){\circle{1.8}}
\put(   6.67,   98.25){\circle{1.8}}
\put(  10.00,   97.68){\circle{1.8}}
\put(  13.33,   96.95){\circle{1.8}}
\put(  16.67,   96.01){\circle{1.8}}
\put(  20.00,   94.85){\circle{1.8}}
\put(  23.33,   93.40){\circle{1.8}}
\put(  26.67,   91.65){\circle{1.8}}
\put(  30.00,   89.56){\circle{1.8}}
\put(  33.33,   87.12){\circle{1.8}}
\put(  36.67,   84.32){\circle{1.8}}
\put(  40.00,   81.18){\circle{1.8}}
\put(  43.33,   77.71){\circle{1.8}}
\put(  46.67,   73.95){\circle{1.8}}
\put(  50.00,   69.95){\circle{1.8}}
\put(  53.33,   65.75){\circle{1.8}}
\put(  56.67,   61.39){\circle{1.8}}
\put(  60.00,   56.92){\circle{1.8}}
\put(  63.33,   52.37){\circle{1.8}}
\put(  66.67,   47.78){\circle{1.8}}
\put(  70.00,   43.17){\circle{1.8}}
\put(  73.33,   38.55){\circle{1.8}}
\put(  76.67,   33.93){\circle{1.8}}
\put(  80.00,   29.32){\circle{1.8}}
\put(  83.33,   24.73){\circle{1.8}}
\put(  86.67,   20.15){\circle{1.8}}
\put(  90.00,   15.60){\circle{1.8}}
\put(  93.33,   11.05){\circle{1.8}}
\put(  96.67,    6.52){\circle{1.8}}

\put(  27.33,   99.97){\circle*{0.8}}
\put(  27.67,   99.52){\circle*{0.8}}
\put(  28.00,   99.07){\circle*{0.8}}
\put(  28.33,   98.62){\circle*{0.8}}
\put(  28.67,   98.18){\circle*{0.8}}
\put(  29.00,   97.73){\circle*{0.8}}
\put(  29.33,   97.28){\circle*{0.8}}
\put(  29.67,   96.84){\circle*{0.8}}
\put(  30.00,   96.39){\circle*{0.8}}
\put(  30.33,   95.94){\circle*{0.8}}
\put(  30.67,   95.50){\circle*{0.8}}
\put(  31.00,   95.05){\circle*{0.8}}
\put(  31.33,   94.60){\circle*{0.8}}
\put(  31.67,   94.16){\circle*{0.8}}
\put(  32.00,   93.71){\circle*{0.8}}
\put(  32.33,   93.26){\circle*{0.8}}
\put(  32.67,   92.82){\circle*{0.8}}
\put(  33.00,   92.37){\circle*{0.8}}
\put(  33.33,   91.92){\circle*{0.8}}
\put(  33.67,   91.47){\circle*{0.8}}
\put(  34.00,   91.03){\circle*{0.8}}
\put(  34.33,   90.58){\circle*{0.8}}
\put(  34.67,   90.13){\circle*{0.8}}
\put(  35.00,   89.69){\circle*{0.8}}
\put(  35.33,   89.24){\circle*{0.8}}
\put(  35.67,   88.79){\circle*{0.8}}
\put(  36.00,   88.35){\circle*{0.8}}
\put(  36.33,   87.90){\circle*{0.8}}
\put(  36.67,   87.45){\circle*{0.8}}
\put(  37.00,   87.01){\circle*{0.8}}
\put(  37.33,   86.56){\circle*{0.8}}
\put(  37.67,   86.11){\circle*{0.8}}
\put(  38.00,   85.67){\circle*{0.8}}
\put(  38.33,   85.22){\circle*{0.8}}
\put(  38.67,   84.77){\circle*{0.8}}
\put(  39.00,   84.32){\circle*{0.8}}
\put(  39.33,   83.88){\circle*{0.8}}
\put(  39.67,   83.43){\circle*{0.8}}
\put(  40.00,   82.98){\circle*{0.8}}
\put(  40.33,   82.54){\circle*{0.8}}
\put(  40.67,   82.09){\circle*{0.8}}
\put(  41.00,   81.64){\circle*{0.8}}
\put(  41.33,   81.20){\circle*{0.8}}
\put(  41.67,   80.75){\circle*{0.8}}
\put(  42.00,   80.30){\circle*{0.8}}
\put(  42.33,   79.86){\circle*{0.8}}
\put(  42.67,   79.41){\circle*{0.8}}
\put(  43.00,   78.96){\circle*{0.8}}
\put(  43.33,   78.52){\circle*{0.8}}
\put(  43.67,   78.07){\circle*{0.8}}
\put(  44.00,   77.62){\circle*{0.8}}
\put(  44.33,   77.18){\circle*{0.8}}
\put(  44.67,   76.73){\circle*{0.8}}
\put(  45.00,   76.28){\circle*{0.8}}
\put(  45.33,   75.83){\circle*{0.8}}
\put(  45.67,   75.39){\circle*{0.8}}
\put(  46.00,   74.94){\circle*{0.8}}
\put(  46.33,   74.49){\circle*{0.8}}
\put(  46.67,   74.05){\circle*{0.8}}
\put(  47.00,   73.60){\circle*{0.8}}
\put(  47.33,   73.15){\circle*{0.8}}
\put(  47.67,   72.71){\circle*{0.8}}
\put(  48.00,   72.26){\circle*{0.8}}
\put(  48.33,   71.81){\circle*{0.8}}
\put(  48.67,   71.37){\circle*{0.8}}
\put(  49.00,   70.92){\circle*{0.8}}
\put(  49.33,   70.47){\circle*{0.8}}
\put(  49.67,   70.03){\circle*{0.8}}
\put(  50.00,   69.58){\circle*{0.8}}
\put(  50.33,   69.13){\circle*{0.8}}
\put(  50.67,   68.68){\circle*{0.8}}
\put(  51.00,   68.24){\circle*{0.8}}
\put(  51.33,   67.79){\circle*{0.8}}
\put(  51.67,   67.34){\circle*{0.8}}
\put(  52.00,   66.90){\circle*{0.8}}
\put(  52.33,   66.45){\circle*{0.8}}
\put(  52.67,   66.00){\circle*{0.8}}
\put(  53.00,   65.56){\circle*{0.8}}
\put(  53.33,   65.11){\circle*{0.8}}
\put(  53.67,   64.66){\circle*{0.8}}
\put(  54.00,   64.22){\circle*{0.8}}
\put(  54.33,   63.77){\circle*{0.8}}
\put(  54.67,   63.32){\circle*{0.8}}
\put(  55.00,   62.88){\circle*{0.8}}
\put(  55.33,   62.43){\circle*{0.8}}
\put(  55.67,   61.98){\circle*{0.8}}
\put(  56.00,   61.54){\circle*{0.8}}
\put(  56.33,   61.09){\circle*{0.8}}
\put(  56.67,   60.64){\circle*{0.8}}
\put(  57.00,   60.19){\circle*{0.8}}
\put(  57.33,   59.75){\circle*{0.8}}
\put(  57.67,   59.30){\circle*{0.8}}
\put(  58.00,   58.85){\circle*{0.8}}
\put(  58.33,   58.41){\circle*{0.8}}
\put(  58.67,   57.96){\circle*{0.8}}
\put(  59.00,   57.51){\circle*{0.8}}
\put(  59.33,   57.07){\circle*{0.8}}
\put(  59.67,   56.62){\circle*{0.8}}
\put(  60.00,   56.17){\circle*{0.8}}
\put(  60.33,   55.73){\circle*{0.8}}
\put(  60.67,   55.28){\circle*{0.8}}
\put(  61.00,   54.83){\circle*{0.8}}
\put(  61.33,   54.39){\circle*{0.8}}
\put(  61.67,   53.94){\circle*{0.8}}
\put(  62.00,   53.49){\circle*{0.8}}
\put(  62.33,   53.04){\circle*{0.8}}
\put(  62.67,   52.60){\circle*{0.8}}
\put(  63.00,   52.15){\circle*{0.8}}
\put(  63.33,   51.70){\circle*{0.8}}
\put(  63.67,   51.26){\circle*{0.8}}
\put(  64.00,   50.81){\circle*{0.8}}
\put(  64.33,   50.36){\circle*{0.8}}
\put(  64.67,   49.92){\circle*{0.8}}
\put(  65.00,   49.47){\circle*{0.8}}
\put(  65.33,   49.02){\circle*{0.8}}
\put(  65.67,   48.58){\circle*{0.8}}
\put(  66.00,   48.13){\circle*{0.8}}
\put(  66.33,   47.68){\circle*{0.8}}
\put(  66.67,   47.24){\circle*{0.8}}
\put(  67.00,   46.79){\circle*{0.8}}
\put(  67.33,   46.34){\circle*{0.8}}
\put(  67.67,   45.89){\circle*{0.8}}
\put(  68.00,   45.45){\circle*{0.8}}
\put(  68.33,   45.00){\circle*{0.8}}
\put(  68.67,   44.55){\circle*{0.8}}
\put(  69.00,   44.11){\circle*{0.8}}
\put(  69.33,   43.66){\circle*{0.8}}
\put(  69.67,   43.21){\circle*{0.8}}
\put(  70.00,   42.77){\circle*{0.8}}
\put(  70.33,   42.32){\circle*{0.8}}
\put(  70.67,   41.87){\circle*{0.8}}
\put(  71.00,   41.43){\circle*{0.8}}
\put(  71.33,   40.98){\circle*{0.8}}
\put(  71.67,   40.53){\circle*{0.8}}
\put(  72.00,   40.09){\circle*{0.8}}
\put(  72.33,   39.64){\circle*{0.8}}
\put(  72.67,   39.19){\circle*{0.8}}
\put(  73.00,   38.75){\circle*{0.8}}
\put(  73.33,   38.30){\circle*{0.8}}
\put(  73.67,   37.85){\circle*{0.8}}
\put(  74.00,   37.40){\circle*{0.8}}
\put(  74.33,   36.96){\circle*{0.8}}
\put(  74.67,   36.51){\circle*{0.8}}
\put(  75.00,   36.06){\circle*{0.8}}
\put(  75.33,   35.62){\circle*{0.8}}
\put(  75.67,   35.17){\circle*{0.8}}
\put(  76.00,   34.72){\circle*{0.8}}
\put(  76.33,   34.28){\circle*{0.8}}
\put(  76.67,   33.83){\circle*{0.8}}
\put(  77.00,   33.38){\circle*{0.8}}
\put(  77.33,   32.94){\circle*{0.8}}
\put(  77.67,   32.49){\circle*{0.8}}
\put(  78.00,   32.04){\circle*{0.8}}
\put(  78.33,   31.60){\circle*{0.8}}
\put(  78.67,   31.15){\circle*{0.8}}
\put(  79.00,   30.70){\circle*{0.8}}
\put(  79.33,   30.25){\circle*{0.8}}
\put(  79.67,   29.81){\circle*{0.8}}
\put(  80.00,   29.36){\circle*{0.8}}
\put(  80.33,   28.91){\circle*{0.8}}
\put(  80.67,   28.47){\circle*{0.8}}
\put(  81.00,   28.02){\circle*{0.8}}
\put(  81.33,   27.57){\circle*{0.8}}
\put(  81.67,   27.13){\circle*{0.8}}
\put(  82.00,   26.68){\circle*{0.8}}
\put(  82.33,   26.23){\circle*{0.8}}
\put(  82.67,   25.79){\circle*{0.8}}
\put(  83.00,   25.34){\circle*{0.8}}
\put(  83.33,   24.89){\circle*{0.8}}
\put(  83.67,   24.45){\circle*{0.8}}
\put(  84.00,   24.00){\circle*{0.8}}
\put(  84.33,   23.55){\circle*{0.8}}
\put(  84.67,   23.10){\circle*{0.8}}
\put(  85.00,   22.66){\circle*{0.8}}
\put(  85.33,   22.21){\circle*{0.8}}
\put(  85.67,   21.76){\circle*{0.8}}
\put(  86.00,   21.32){\circle*{0.8}}
\put(  86.33,   20.87){\circle*{0.8}}
\put(  86.67,   20.42){\circle*{0.8}}
\put(  87.00,   19.98){\circle*{0.8}}
\put(  87.33,   19.53){\circle*{0.8}}
\put(  87.67,   19.08){\circle*{0.8}}
\put(  88.00,   18.64){\circle*{0.8}}
\put(  88.33,   18.19){\circle*{0.8}}
\put(  88.67,   17.74){\circle*{0.8}}
\put(  89.00,   17.30){\circle*{0.8}}
\put(  89.33,   16.85){\circle*{0.8}}
\put(  89.67,   16.40){\circle*{0.8}}
\put(  90.00,   15.96){\circle*{0.8}}
\put(  90.33,   15.51){\circle*{0.8}}
\put(  90.67,   15.06){\circle*{0.8}}
\put(  91.00,   14.61){\circle*{0.8}}
\put(  91.33,   14.17){\circle*{0.8}}
\put(  91.67,   13.72){\circle*{0.8}}
\put(  92.00,   13.27){\circle*{0.8}}
\put(  92.33,   12.83){\circle*{0.8}}
\put(  92.67,   12.38){\circle*{0.8}}
\put(  93.00,   11.93){\circle*{0.8}}
\put(  93.33,   11.49){\circle*{0.8}}
\put(  93.67,   11.04){\circle*{0.8}}
\put(  94.00,   10.59){\circle*{0.8}}
\put(  94.33,   10.15){\circle*{0.8}}
\put(  94.67,    9.70){\circle*{0.8}}
\put(  95.00,    9.25){\circle*{0.8}}
\put(  95.33,    8.81){\circle*{0.8}}
\put(  95.67,    8.36){\circle*{0.8}}
\put(  96.00,    7.91){\circle*{0.8}}
\put(  96.33,    7.46){\circle*{0.8}}
\put(  96.67,    7.02){\circle*{0.8}}
\put(  97.00,    6.57){\circle*{0.8}}
\put(  97.33,    6.12){\circle*{0.8}}
\put(  97.67,    5.68){\circle*{0.8}}
\put(  98.00,    5.23){\circle*{0.8}}
\put(  98.33,    4.78){\circle*{0.8}}
\put(  98.67,    4.34){\circle*{0.8}}
\put(  99.00,    3.89){\circle*{0.8}}
\put(  99.33,    3.44){\circle*{0.8}}
\put(  99.67,    3.00){\circle*{0.8}}

\end{picture}
\end{center}
\vspace*{10 mm}

\caption{}
\end{figure}

\setlength{\unitlength}{0.75 mm}
\begin{figure}[tbh]
\begin{center}

\end{center}
\vspace*{10 mm}

\caption{}
\end{figure}

\setlength{\unitlength}{0.75 mm}
\begin{figure}[tbh]
\begin{center}
%
\end{center}
\vspace*{10 mm}

\caption{}
\end{figure}

\setlength{\unitlength}{0.75 mm}
\begin{figure}[tbh]
\begin{center}
%
\end{center}
\vspace*{10 mm}

\caption{}
\end{figure}

\setlength{\unitlength}{0.75 mm}
\begin{figure}[tbh]
\begin{center}
\begin{picture}(100,100)
\put(0,0){\framebox(100,100)}
\multiput(12.5,0)(12.5,0){7}{\line(0,1){2}}
\multiput(0,16.67)(0,16.67){5}{\line(1,0){2}}
\put(0,-9){$-2.0$}
\put(94,-9){2.0}
\put(50,-9){$\log \zeta$}
\put(-16,0){$-12.0$}
\put(-10,96){0.0}
\put(-15,70){$\log \overline{S}$}
\put(49,103){1}
\put(31,103){2}
\put(36.8,0){\line(0,1){100}}

\put(   3.33,   76.70){\circle{1.8}}
\put(   6.67,   78.36){\circle{1.8}}
\put(  10.00,   80.02){\circle{1.8}}
\put(  13.33,   81.68){\circle{1.8}}
\put(  16.67,   83.34){\circle{1.8}}
\put(  20.00,   85.00){\circle{1.8}}
\put(  23.33,   86.67){\circle{1.8}}
\put(  26.67,   88.33){\circle{1.8}}
\put(  30.00,   89.65){\circle{1.8}}
\put(  33.33,   90.01){\circle{1.8}}
\put(  36.67,   89.04){\circle{1.8}}
\put(  40.00,   86.82){\circle{1.8}}
\put(  43.33,   83.69){\circle{1.8}}
\put(  46.67,   80.00){\circle{1.8}}
\put(  50.00,   75.97){\circle{1.8}}
\put(  53.33,   71.76){\circle{1.8}}
\put(  56.67,   67.44){\circle{1.8}}
\put(  60.00,   63.06){\circle{1.8}}
\put(  63.33,   58.65){\circle{1.8}}
\put(  66.67,   54.23){\circle{1.8}}
\put(  70.00,   49.79){\circle{1.8}}
\put(  73.33,   45.36){\circle{1.8}}
\put(  76.67,   40.91){\circle{1.8}}
\put(  80.00,   36.47){\circle{1.8}}
\put(  83.33,   32.03){\circle{1.8}}
\put(  86.67,   27.58){\circle{1.8}}
\put(  90.00,   23.14){\circle{1.8}}
\put(  93.33,   18.70){\circle{1.8}}
\put(  96.67,   14.25){\circle{1.8}}
\put( 100.00,    9.81){\circle{1.8}}

\put(   0.33,   75.28){\circle*{0.8}}
\put(   0.67,   75.44){\circle*{0.8}}
\put(   1.00,   75.60){\circle*{0.8}}
\put(   1.33,   75.77){\circle*{0.8}}
\put(   1.67,   75.93){\circle*{0.8}}
\put(   2.00,   76.10){\circle*{0.8}}
\put(   2.33,   76.26){\circle*{0.8}}
\put(   2.67,   76.42){\circle*{0.8}}
\put(   3.00,   76.59){\circle*{0.8}}
\put(   3.33,   76.75){\circle*{0.8}}
\put(   3.67,   76.92){\circle*{0.8}}
\put(   4.00,   77.08){\circle*{0.8}}
\put(   4.33,   77.24){\circle*{0.8}}
\put(   4.67,   77.41){\circle*{0.8}}
\put(   5.00,   77.57){\circle*{0.8}}
\put(   5.33,   77.73){\circle*{0.8}}
\put(   5.67,   77.90){\circle*{0.8}}
\put(   6.00,   78.06){\circle*{0.8}}
\put(   6.33,   78.23){\circle*{0.8}}
\put(   6.67,   78.39){\circle*{0.8}}
\put(   7.00,   78.55){\circle*{0.8}}
\put(   7.33,   78.72){\circle*{0.8}}
\put(   7.67,   78.88){\circle*{0.8}}
\put(   8.00,   79.05){\circle*{0.8}}
\put(   8.33,   79.21){\circle*{0.8}}
\put(   8.67,   79.37){\circle*{0.8}}
\put(   9.00,   79.54){\circle*{0.8}}
\put(   9.33,   79.70){\circle*{0.8}}
\put(   9.67,   79.86){\circle*{0.8}}
\put(  10.00,   80.03){\circle*{0.8}}
\put(  10.33,   80.19){\circle*{0.8}}
\put(  10.67,   80.36){\circle*{0.8}}
\put(  11.00,   80.52){\circle*{0.8}}
\put(  11.33,   80.68){\circle*{0.8}}
\put(  11.67,   80.85){\circle*{0.8}}
\put(  12.00,   81.01){\circle*{0.8}}
\put(  12.33,   81.18){\circle*{0.8}}
\put(  12.67,   81.34){\circle*{0.8}}
\put(  13.00,   81.50){\circle*{0.8}}
\put(  13.33,   81.67){\circle*{0.8}}
\put(  13.67,   81.83){\circle*{0.8}}
\put(  14.00,   82.00){\circle*{0.8}}
\put(  14.33,   82.16){\circle*{0.8}}
\put(  14.67,   82.32){\circle*{0.8}}
\put(  15.00,   82.49){\circle*{0.8}}
\put(  15.33,   82.65){\circle*{0.8}}
\put(  15.67,   82.81){\circle*{0.8}}
\put(  16.00,   82.98){\circle*{0.8}}
\put(  16.33,   83.14){\circle*{0.8}}
\put(  16.67,   83.31){\circle*{0.8}}
\put(  17.00,   83.47){\circle*{0.8}}
\put(  17.33,   83.63){\circle*{0.8}}
\put(  17.67,   83.80){\circle*{0.8}}
\put(  18.00,   83.96){\circle*{0.8}}
\put(  18.33,   84.13){\circle*{0.8}}
\put(  18.67,   84.29){\circle*{0.8}}
\put(  19.00,   84.45){\circle*{0.8}}
\put(  19.33,   84.62){\circle*{0.8}}
\put(  19.67,   84.78){\circle*{0.8}}
\put(  20.00,   84.94){\circle*{0.8}}
\put(  20.33,   85.11){\circle*{0.8}}
\put(  20.67,   85.27){\circle*{0.8}}
\put(  21.00,   85.44){\circle*{0.8}}
\put(  21.33,   85.60){\circle*{0.8}}
\put(  21.67,   85.76){\circle*{0.8}}
\put(  22.00,   85.93){\circle*{0.8}}
\put(  22.33,   86.09){\circle*{0.8}}
\put(  22.67,   86.26){\circle*{0.8}}
\put(  23.00,   86.42){\circle*{0.8}}
\put(  23.33,   86.58){\circle*{0.8}}
\put(  23.67,   86.75){\circle*{0.8}}
\put(  24.00,   86.91){\circle*{0.8}}
\put(  24.33,   87.08){\circle*{0.8}}
\put(  24.67,   87.24){\circle*{0.8}}
\put(  25.00,   87.40){\circle*{0.8}}
\put(  25.33,   87.57){\circle*{0.8}}
\put(  25.67,   87.73){\circle*{0.8}}
\put(  26.00,   87.89){\circle*{0.8}}
\put(  26.33,   88.06){\circle*{0.8}}
\put(  26.67,   88.22){\circle*{0.8}}
\put(  27.00,   88.39){\circle*{0.8}}
\put(  27.33,   88.55){\circle*{0.8}}
\put(  27.67,   88.71){\circle*{0.8}}
\put(  28.00,   88.88){\circle*{0.8}}
\put(  28.33,   89.04){\circle*{0.8}}
\put(  28.67,   89.21){\circle*{0.8}}
\put(  29.00,   89.37){\circle*{0.8}}
\put(  29.33,   89.53){\circle*{0.8}}
\put(  29.67,   89.70){\circle*{0.8}}
\put(  30.00,   89.86){\circle*{0.8}}
\put(  30.33,   90.02){\circle*{0.8}}
\put(  30.67,   90.19){\circle*{0.8}}
\put(  31.00,   90.35){\circle*{0.8}}
\put(  31.33,   90.52){\circle*{0.8}}
\put(  31.67,   90.68){\circle*{0.8}}
\put(  32.00,   90.84){\circle*{0.8}}
\put(  32.33,   91.01){\circle*{0.8}}
\put(  32.67,   91.17){\circle*{0.8}}
\put(  33.00,   91.34){\circle*{0.8}}
\put(  33.33,   91.50){\circle*{0.8}}
\put(  33.67,   91.66){\circle*{0.8}}
\put(  34.00,   91.83){\circle*{0.8}}
\put(  34.33,   91.99){\circle*{0.8}}
\put(  34.67,   92.16){\circle*{0.8}}
\put(  35.00,   92.32){\circle*{0.8}}
\put(  35.33,   92.48){\circle*{0.8}}
\put(  35.67,   92.65){\circle*{0.8}}
\put(  36.00,   92.81){\circle*{0.8}}
\put(  36.33,   92.97){\circle*{0.8}}
\put(  36.67,   93.14){\circle*{0.8}}
\put(  37.00,   93.30){\circle*{0.8}}
\put(  37.33,   93.47){\circle*{0.8}}
\put(  37.67,   93.63){\circle*{0.8}}
\put(  38.00,   93.79){\circle*{0.8}}
\put(  38.33,   93.96){\circle*{0.8}}
\put(  38.67,   94.12){\circle*{0.8}}
\put(  39.00,   94.29){\circle*{0.8}}
\put(  39.33,   94.45){\circle*{0.8}}
\put(  39.67,   94.61){\circle*{0.8}}
\put(  40.00,   94.78){\circle*{0.8}}
\put(  40.33,   94.94){\circle*{0.8}}
\put(  40.67,   95.10){\circle*{0.8}}
\put(  41.00,   95.27){\circle*{0.8}}
\put(  41.33,   95.43){\circle*{0.8}}
\put(  41.67,   95.60){\circle*{0.8}}
\put(  42.00,   95.76){\circle*{0.8}}
\put(  42.33,   95.92){\circle*{0.8}}
\put(  42.67,   96.09){\circle*{0.8}}
\put(  43.00,   96.25){\circle*{0.8}}
\put(  43.33,   96.42){\circle*{0.8}}
\put(  43.67,   96.58){\circle*{0.8}}
\put(  44.00,   96.74){\circle*{0.8}}
\put(  44.33,   96.91){\circle*{0.8}}
\put(  44.67,   97.07){\circle*{0.8}}
\put(  45.00,   97.24){\circle*{0.8}}
\put(  45.33,   97.40){\circle*{0.8}}
\put(  45.67,   97.56){\circle*{0.8}}
\put(  46.00,   97.73){\circle*{0.8}}
\put(  46.33,   97.89){\circle*{0.8}}
\put(  46.67,   98.05){\circle*{0.8}}
\put(  47.00,   98.22){\circle*{0.8}}
\put(  47.33,   98.38){\circle*{0.8}}
\put(  47.67,   98.55){\circle*{0.8}}
\put(  48.00,   98.71){\circle*{0.8}}
\put(  48.33,   98.87){\circle*{0.8}}
\put(  48.67,   99.04){\circle*{0.8}}
\put(  49.00,   99.20){\circle*{0.8}}
\put(  49.33,   99.37){\circle*{0.8}}
\put(  49.67,   99.53){\circle*{0.8}}
\put(  50.00,   99.69){\circle*{0.8}}
\put(  50.33,   99.86){\circle*{0.8}}

\put(  32.00,   99.94){\circle*{0.8}}
\put(  32.33,   99.50){\circle*{0.8}}
\put(  32.67,   99.06){\circle*{0.8}}
\put(  33.00,   98.62){\circle*{0.8}}
\put(  33.33,   98.17){\circle*{0.8}}
\put(  33.67,   97.73){\circle*{0.8}}
\put(  34.00,   97.29){\circle*{0.8}}
\put(  34.33,   96.85){\circle*{0.8}}
\put(  34.67,   96.41){\circle*{0.8}}
\put(  35.00,   95.97){\circle*{0.8}}
\put(  35.33,   95.53){\circle*{0.8}}
\put(  35.67,   95.09){\circle*{0.8}}
\put(  36.00,   94.65){\circle*{0.8}}
\put(  36.33,   94.20){\circle*{0.8}}
\put(  36.67,   93.76){\circle*{0.8}}
\put(  37.00,   93.32){\circle*{0.8}}
\put(  37.33,   92.88){\circle*{0.8}}
\put(  37.67,   92.44){\circle*{0.8}}
\put(  38.00,   92.00){\circle*{0.8}}
\put(  38.33,   91.56){\circle*{0.8}}
\put(  38.67,   91.12){\circle*{0.8}}
\put(  39.00,   90.68){\circle*{0.8}}
\put(  39.33,   90.23){\circle*{0.8}}
\put(  39.67,   89.79){\circle*{0.8}}
\put(  40.00,   89.35){\circle*{0.8}}
\put(  40.33,   88.91){\circle*{0.8}}
\put(  40.67,   88.47){\circle*{0.8}}
\put(  41.00,   88.03){\circle*{0.8}}
\put(  41.33,   87.59){\circle*{0.8}}
\put(  41.67,   87.15){\circle*{0.8}}
\put(  42.00,   86.71){\circle*{0.8}}
\put(  42.33,   86.26){\circle*{0.8}}
\put(  42.67,   85.82){\circle*{0.8}}
\put(  43.00,   85.38){\circle*{0.8}}
\put(  43.33,   84.94){\circle*{0.8}}
\put(  43.67,   84.50){\circle*{0.8}}
\put(  44.00,   84.06){\circle*{0.8}}
\put(  44.33,   83.62){\circle*{0.8}}
\put(  44.67,   83.18){\circle*{0.8}}
\put(  45.00,   82.74){\circle*{0.8}}
\put(  45.33,   82.30){\circle*{0.8}}
\put(  45.67,   81.85){\circle*{0.8}}
\put(  46.00,   81.41){\circle*{0.8}}
\put(  46.33,   80.97){\circle*{0.8}}
\put(  46.67,   80.53){\circle*{0.8}}
\put(  47.00,   80.09){\circle*{0.8}}
\put(  47.33,   79.65){\circle*{0.8}}
\put(  47.67,   79.21){\circle*{0.8}}
\put(  48.00,   78.77){\circle*{0.8}}
\put(  48.33,   78.33){\circle*{0.8}}
\put(  48.67,   77.88){\circle*{0.8}}
\put(  49.00,   77.44){\circle*{0.8}}
\put(  49.33,   77.00){\circle*{0.8}}
\put(  49.67,   76.56){\circle*{0.8}}
\put(  50.00,   76.12){\circle*{0.8}}
\put(  50.33,   75.68){\circle*{0.8}}
\put(  50.67,   75.24){\circle*{0.8}}
\put(  51.00,   74.80){\circle*{0.8}}
\put(  51.33,   74.36){\circle*{0.8}}
\put(  51.67,   73.91){\circle*{0.8}}
\put(  52.00,   73.47){\circle*{0.8}}
\put(  52.33,   73.03){\circle*{0.8}}
\put(  52.67,   72.59){\circle*{0.8}}
\put(  53.00,   72.15){\circle*{0.8}}
\put(  53.33,   71.71){\circle*{0.8}}
\put(  53.67,   71.27){\circle*{0.8}}
\put(  54.00,   70.83){\circle*{0.8}}
\put(  54.33,   70.39){\circle*{0.8}}
\put(  54.67,   69.94){\circle*{0.8}}
\put(  55.00,   69.50){\circle*{0.8}}
\put(  55.33,   69.06){\circle*{0.8}}
\put(  55.67,   68.62){\circle*{0.8}}
\put(  56.00,   68.18){\circle*{0.8}}
\put(  56.33,   67.74){\circle*{0.8}}
\put(  56.67,   67.30){\circle*{0.8}}
\put(  57.00,   66.86){\circle*{0.8}}
\put(  57.33,   66.42){\circle*{0.8}}
\put(  57.67,   65.98){\circle*{0.8}}
\put(  58.00,   65.53){\circle*{0.8}}
\put(  58.33,   65.09){\circle*{0.8}}
\put(  58.67,   64.65){\circle*{0.8}}
\put(  59.00,   64.21){\circle*{0.8}}
\put(  59.33,   63.77){\circle*{0.8}}
\put(  59.67,   63.33){\circle*{0.8}}
\put(  60.00,   62.89){\circle*{0.8}}
\put(  60.33,   62.45){\circle*{0.8}}
\put(  60.67,   62.01){\circle*{0.8}}
\put(  61.00,   61.56){\circle*{0.8}}
\put(  61.33,   61.12){\circle*{0.8}}
\put(  61.67,   60.68){\circle*{0.8}}
\put(  62.00,   60.24){\circle*{0.8}}
\put(  62.33,   59.80){\circle*{0.8}}
\put(  62.67,   59.36){\circle*{0.8}}
\put(  63.00,   58.92){\circle*{0.8}}
\put(  63.33,   58.48){\circle*{0.8}}
\put(  63.67,   58.04){\circle*{0.8}}
\put(  64.00,   57.59){\circle*{0.8}}
\put(  64.33,   57.15){\circle*{0.8}}
\put(  64.67,   56.71){\circle*{0.8}}
\put(  65.00,   56.27){\circle*{0.8}}
\put(  65.33,   55.83){\circle*{0.8}}
\put(  65.67,   55.39){\circle*{0.8}}
\put(  66.00,   54.95){\circle*{0.8}}
\put(  66.33,   54.51){\circle*{0.8}}
\put(  66.67,   54.07){\circle*{0.8}}
\put(  67.00,   53.62){\circle*{0.8}}
\put(  67.33,   53.18){\circle*{0.8}}
\put(  67.67,   52.74){\circle*{0.8}}
\put(  68.00,   52.30){\circle*{0.8}}
\put(  68.33,   51.86){\circle*{0.8}}
\put(  68.67,   51.42){\circle*{0.8}}
\put(  69.00,   50.98){\circle*{0.8}}
\put(  69.33,   50.54){\circle*{0.8}}
\put(  69.67,   50.10){\circle*{0.8}}
\put(  70.00,   49.66){\circle*{0.8}}
\put(  70.33,   49.21){\circle*{0.8}}
\put(  70.67,   48.77){\circle*{0.8}}
\put(  71.00,   48.33){\circle*{0.8}}
\put(  71.33,   47.89){\circle*{0.8}}
\put(  71.67,   47.45){\circle*{0.8}}
\put(  72.00,   47.01){\circle*{0.8}}
\put(  72.33,   46.57){\circle*{0.8}}
\put(  72.67,   46.13){\circle*{0.8}}
\put(  73.00,   45.69){\circle*{0.8}}
\put(  73.33,   45.24){\circle*{0.8}}
\put(  73.67,   44.80){\circle*{0.8}}
\put(  74.00,   44.36){\circle*{0.8}}
\put(  74.33,   43.92){\circle*{0.8}}
\put(  74.67,   43.48){\circle*{0.8}}
\put(  75.00,   43.04){\circle*{0.8}}
\put(  75.33,   42.60){\circle*{0.8}}
\put(  75.67,   42.16){\circle*{0.8}}
\put(  76.00,   41.72){\circle*{0.8}}
\put(  76.33,   41.27){\circle*{0.8}}
\put(  76.67,   40.83){\circle*{0.8}}
\put(  77.00,   40.39){\circle*{0.8}}
\put(  77.33,   39.95){\circle*{0.8}}
\put(  77.67,   39.51){\circle*{0.8}}
\put(  78.00,   39.07){\circle*{0.8}}
\put(  78.33,   38.63){\circle*{0.8}}
\put(  78.67,   38.19){\circle*{0.8}}
\put(  79.00,   37.75){\circle*{0.8}}
\put(  79.33,   37.30){\circle*{0.8}}
\put(  79.67,   36.86){\circle*{0.8}}
\put(  80.00,   36.42){\circle*{0.8}}
\put(  80.33,   35.98){\circle*{0.8}}
\put(  80.67,   35.54){\circle*{0.8}}
\put(  81.00,   35.10){\circle*{0.8}}
\put(  81.33,   34.66){\circle*{0.8}}
\put(  81.67,   34.22){\circle*{0.8}}
\put(  82.00,   33.78){\circle*{0.8}}
\put(  82.33,   33.34){\circle*{0.8}}
\put(  82.67,   32.89){\circle*{0.8}}
\put(  83.00,   32.45){\circle*{0.8}}
\put(  83.33,   32.01){\circle*{0.8}}
\put(  83.67,   31.57){\circle*{0.8}}
\put(  84.00,   31.13){\circle*{0.8}}
\put(  84.33,   30.69){\circle*{0.8}}
\put(  84.67,   30.25){\circle*{0.8}}
\put(  85.00,   29.81){\circle*{0.8}}
\put(  85.33,   29.37){\circle*{0.8}}
\put(  85.67,   28.92){\circle*{0.8}}
\put(  86.00,   28.48){\circle*{0.8}}
\put(  86.33,   28.04){\circle*{0.8}}
\put(  86.67,   27.60){\circle*{0.8}}
\put(  87.00,   27.16){\circle*{0.8}}
\put(  87.33,   26.72){\circle*{0.8}}
\put(  87.67,   26.28){\circle*{0.8}}
\put(  88.00,   25.84){\circle*{0.8}}
\put(  88.33,   25.40){\circle*{0.8}}
\put(  88.67,   24.95){\circle*{0.8}}
\put(  89.00,   24.51){\circle*{0.8}}
\put(  89.33,   24.07){\circle*{0.8}}
\put(  89.67,   23.63){\circle*{0.8}}
\put(  90.00,   23.19){\circle*{0.8}}
\put(  90.33,   22.75){\circle*{0.8}}
\put(  90.67,   22.31){\circle*{0.8}}
\put(  91.00,   21.87){\circle*{0.8}}
\put(  91.33,   21.43){\circle*{0.8}}
\put(  91.67,   20.98){\circle*{0.8}}
\put(  92.00,   20.54){\circle*{0.8}}
\put(  92.33,   20.10){\circle*{0.8}}
\put(  92.67,   19.66){\circle*{0.8}}
\put(  93.00,   19.22){\circle*{0.8}}
\put(  93.33,   18.78){\circle*{0.8}}
\put(  93.67,   18.34){\circle*{0.8}}
\put(  94.00,   17.90){\circle*{0.8}}
\put(  94.33,   17.46){\circle*{0.8}}
\put(  94.67,   17.02){\circle*{0.8}}
\put(  95.00,   16.57){\circle*{0.8}}
\put(  95.33,   16.13){\circle*{0.8}}
\put(  95.67,   15.69){\circle*{0.8}}
\put(  96.00,   15.25){\circle*{0.8}}
\put(  96.33,   14.81){\circle*{0.8}}
\put(  96.67,   14.37){\circle*{0.8}}
\put(  97.00,   13.93){\circle*{0.8}}
\put(  97.33,   13.49){\circle*{0.8}}
\put(  97.67,   13.05){\circle*{0.8}}
\put(  98.00,   12.60){\circle*{0.8}}
\put(  98.33,   12.16){\circle*{0.8}}
\put(  98.67,   11.72){\circle*{0.8}}
\put(  99.00,   11.28){\circle*{0.8}}
\put(  99.33,   10.84){\circle*{0.8}}
\put(  99.67,   10.40){\circle*{0.8}}

\end{picture}
\end{center}
\vspace*{10 mm}

\caption{}
\end{figure}
\end{document}